\RequirePackage{fix-cm}
\documentclass[smallcondensed]{svjour3} 
\smartqed 
\usepackage{graphicx}
\usepackage{epstopdf}
\usepackage{mathptmx} 
\usepackage{amsmath,amssymb,color,enumerate,latexsym,multirow,natbib,placeins}
\begin{document}

\title{Hilbert transform, spectral filters and option pricing
\thanks{The support of the Economic and Social Research Council (ESRC) in funding the Systemic Risk Centre (grant number ES/K002309/1) and of the Engineering and Physical Sciences Research Council (EPSRC) in funding the UK Centre for Doctoral Training in Financial Computing and Analytics (grant number 1482817) are gratefully acknowledged.}
}


\author{Carolyn E.~Phelan \and Daniele Marazzina \and Gianluca~Fusai \and
Guido Germano}

\authorrunning{Phelan et al.} 

\institute{Carolyn E.~Phelan \at
Department of Computer Science, University College London\\
\email{c.phelan@cs.ucl.ac.uk, carolyn.phelan.14@ucl.ac.uk}
\and Gianluca Fusai \at
Dipartimento di Studi per l'Economia e l'Impresa (DiSEI), Universit\`a del Piemonte Orientale ``Amedeo Avogadro'', Novara \\
Faculty of Finance, Cass Business School, City, University London \\
\email{gianluca.fusai@uniupo.it, gianluca.fusai.1@city.ac.uk}
\and Daniele Marazzina \at
Dipartimento di Matematica, Politecnico di Milano \\
\email{daniele.marazzina@polimi.it}
\and Guido Germano \at
Department of Computer Science, University College London \\
Systemic Risk Centre, London School of Economics and Political Science \\
\email{g.germano@ucl.ac.uk, g.germano@lse.ac.uk}
}

\date{Received: date / Accepted: date}

\maketitle

\begin{abstract}
We show how spectral filters can improve the convergence of numerical schemes which use discrete Hilbert transforms based on a sinc function expansion, and thus ultimately on the fast Fourier transform. This is relevant, for example, for the computation of fluctuation identities, which give the distribution of the maximum or the minimum of a random path, or the joint distribution at maturity with the extrema staying below or above barriers. We use as examples the methods by Feng and Linetsky (2008) and Fusai, Germano and Marazzina (2016) to price discretely monitored barrier options where the underlying asset price is modelled by an exponential L\'evy process. Both methods show exponential convergence with respect to the number of grid points in most cases, but are limited to polynomial convergence under certain conditions. We relate these rates of convergence to the Gibbs phenomenon for Fourier transforms and achieve improved results with spectral filtering.
\keywords{double-barrier option \and discrete monitoring \and L\'evy processes \and Spitzer identity \and Wiener-Hopf factorisation \and Hilbert transform \and Fourier transform \and FFT \and $z$-transform \and sinc function \and Gibbs phenomenon \and spectral filter}
\end{abstract}

\section{Introduction}
Derivative pricing with Fourier transforms was first investigated by \cite{heston1993closed}. \cite{Carr1999} published the first method with both the characteristic function and the payoff in the Fourier domain. \cite{fang2008novel,Fang2009pricing} devised the COS method based on the Fourier-cosine expansion. The Hilbert transform \citep{King2009} has also been successfully employed: by \cite{Feng2008} to price barrier options using backward induction in the Fourier space and by \cite{marazzina2012pricing} and \cite{fusai2015} to compute the factorisations required by the Spitzer identities \citep{spitzer1956combinatorial,Kemperman1963} via the Plemelj-Sokhotski relations. Pricing derivatives, especially exotic options, is a challenging problem in the operations research literature. \cite{fusai2015} provide extensive references for this, as well as for many non-financial applications of the Hilbert transform and the related topics of Wiener-Hopf factorisation and Spitzer identities in insurance, queuing theory, physics, engineering, applied mathematics, etc.

When working in the Fourier domain, the use of numerical solutions means that one must manage issues arising from the approximation of an integral over an infinite domain with a finite sum. As long as the truncation limits in the log-price domain are selected judiciously, the main issue to contend with is the so-called Gibbs phenomenon \citep{wilbraham1848certain,gibbs1898fourier,gibbs1899fourier}. There have been many papers exploring possible solutions to the Gibbs phenomenon in a general setting, notably by \cite{hewitt1979gibbs}, \cite{vandeven1991family}, \cite{gottlieb1997gibbs}, \cite{tadmor2005adaptive} and \cite{tadmor2007filters}. More recently \cite{ruijter2015application} explored the use of spectral filtering techniques to solve the problem of slow polynomial error convergence seen when the COS method is used with non-smooth probability distributions.

The application investigated in this paper is the improvement of methods based on the fast Hilbert transform for the pricing of discretely monitored barrier options with L\'evy processes. Recent papers have described significant progress in the valuation of these types of financial contracts. \cite{Feng2008} devised a method using the sinc-based Hilbert transform by \cite{Stenger1993,Stenger2011} whose convergence is exponential for many L\'evy processes but only polynomial for the variance gamma (VG) process; this method has a computational time which increases linearly with the number of monitoring dates $N$. \cite{fusai2015} used the Spitzer identities to devise a method whose computation time is independent of $N$ and which achieves exponential convergence for single-barrier and lookback options, again with the exception of the VG process, but which is limited to polynomial convergence for double-barrier options.

Our main contribution to the literature on this subject is to produce an error bound explaining the origin of the polynomial error convergence seen by \cite{fusai2015} and then to extend this method to exponential convergence with the use of spectral filters. We have thereby produced the first pricing method for discretely monitored double barrier options whose error converges exponentially on the number of grid points and whose CPU time is independent of the number of monitoring dates. We also show that the pricing method by \cite{Feng2008} can be improved for the VG process and present improved results for both methods with spectral filters. In the course of this work we extend the investigation into the error performance of the sinc-based Hilbert transform and show that the error convergence is related to both the shape of the characteristic function of the underlying process and to the Gibbs phenomenon. We also implement the exponential filter which was first used in pricing applications by \cite{ruijter2015application} and the Planck taper by \cite{mckechan2010tapering}. To our knowledge, this is the first occasion that the latter has been used in financial mathematics.

The structure of this paper is as follows. In Section~\ref{sec:Back} we run briefly through Fourier, Hilbert and $z$-transforms, give a concise overview of the original pricing schemes and explain our modifications to improve convergence. Section~\ref{sec:Errperf} includes a discussion of the performances of the pricing techniques and how they relate to the Gibbs phenomenon and the shape of the characteristic function of the underlying processes. Lastly, Section~\ref{sec:results} shows numerical results, comparing the filtered algorithms with the original methods. Further numerical tests are presented in the online supplementary material.

\section{Background}\label{sec:Back}
As this method directly extends the Fusai, Germano and Marazzina (FGM) \citep{fusai2015} and FL \citep{Feng2008} pricing methods, we refer to the original papers for a comprehensive introduction. Aspects of the methods which are directly relevant to the error investigation are described here in order to provide a background to the changes that we made to improve convergence.
\subsection{Fourier and Hilbert transforms}\label{sec:Back_fourhilb}
In this paper we make extensive use of the Fourier transform \citep[see e.g.][]{Polyanin1998,Kreyszig2011}, an integral transform with many applications. 
Historically, it has been widely used in spectroscopy and communications, therefore much of the literature refers to the function in the Fourier domain as its spectrum. 
According to the usual convention in the finance literature, the forward and inverse Fourier transforms are defined as
\begin{align}
& \widehat{f}(\xi)=\mathcal{F}_{x\rightarrow\xi} \left[f(x)\right]=\int^{+\infty}_{-\infty}f(x)e^{i\xi x}dx, \label{eq:FwdFourier}\\
& f(x)=\mathcal{F}^{-1}_{\xi\rightarrow x} \left[\widehat{f}(\xi)\right]=\frac{1}{2\pi}\int^{+\infty}_{-\infty}\widehat{f}(\xi)e^{-i\xi x}d\xi \label{eq:RevFourier}.
\end{align}
Let $S(t)$ be the price of an underlying asset and $x(t) = \log(S(t)/S_0)$ its log-price, where $S_0 = S(0)$. Let us consider a derivative characterized by a payoff at maturity $T$. E.g., a double-barrier option with strike $K$ and upper (lower) barrier $U$ ($L$) has the damped payoff
\begin{equation}
\phi(x(T)) = e^{\alpha x}S_0\max(\theta(e^x-e^k),0)\mathbf{1}_{(l,u)}(x),\label{eq:dpayoff}
\end{equation}
where $\theta = 1$ for a call, $\theta = -1$ for a put, $\mathbf{1}_A(x)$ is the indicator function of the set $A$, $k=\log(K/S_0)$ is the log-strike, $u=\log(U/S_0)$ is the upper log-barrier and $l=\log(L/S_0)$ is the lower log-barrier. The damping factor $e^{\alpha x}$ is used to ensure the integrability of the payoff function; see \cite{Feng2008} for a full discussion of the selection of the damping parameter $\alpha$. To find the price $v(x,t)$ of an option at time $t=0$ when the initial price of the underlying is $S_0$ and thus its log-price is $x(0)=0$, we need to discount the expected value of the undamped option payoff $\phi(x(T))e^{-\alpha x(T)}$ at maturity $t=T$ with respect to an appropriate risk-neutral probability distribution function (PDF) $p(x,T)$ whose initial condition is $p(x,0) = \delta(x)$. As shown by \cite{lewis2001simple}, this can be done using the Plancherel relation,
\begin{align}
v(0,0) & = e^{-rT}\mathrm{E}\left[\phi(x(T))e^{-\alpha x(T)}|x(0)=0\right]=e^{-rT}\int^{+\infty}_{-\infty}\phi(x)e^{-\alpha x}p(x,T)dx \nonumber\\
& = \frac{e^{-rT}}{2\pi}\int^{+\infty}_{-\infty}\widehat{\phi}(\xi)\widehat{p}\,^*(\xi+i\alpha,T)d\xi = e^{-rT}\mathcal{F}^{-1}_{\xi\rightarrow x}\left[\widehat{\phi}(\xi)\widehat{p}\,^*(\xi+i\alpha,T)\right](0). \label{eq:Planch}
\end{align}
Here, $\widehat{p}\,^*(\xi+i\alpha,T)$ is the complex conjugate of the Fourier transform of $e^{-\alpha x}p(x,T)$. To price options using this relation, we need the Fourier transforms of both the damped payoff and the PDF.
 The Fourier transform of the damped payoff $\phi(x)$ is available analytically, 
\begin{equation}
\label{eq:Payoff}
\widehat{\phi}(\xi)=S_0\left(\frac{e^{(1+i\xi+\alpha)a}-e^{(1+i\xi+\alpha)b}}{1+i\xi+\alpha}-\frac{e^{k+(i\xi+\alpha)a}-e^{k+(i\xi+\alpha)b}}{i\xi+\alpha}\right),
\end{equation}
where for a call option $a = u$ and $b = \max(k,l)$, while for a put option $a=l$ and $b=\min(k,u)$.

The Fourier transform of the PDF $p(x,t)$ of a stochastic process $X(t)$ is the characteristic function
\begin{equation}
\label{eq:CharFun}
\Psi(\xi,t)=\mathrm{E}\left[e^{i\xi X(t)}\right]=\int^{+\infty}_{-\infty}p(x,t) e^{i\xi x}dx=\mathcal{F}_{x\rightarrow\xi}\left[p(x,t)\right]=\widehat{p}(\xi,t).
\end{equation}
For a L\'evy process the characteristic function can be written as $\Psi(\xi,t)=e^{\psi(\xi)t}$, where the characteristic exponent $\psi(\xi)$ is given by the L\'evy-Khinchine formula as
\begin{equation}
\label{eq:CharExp}
\psi(\xi)=i\mu\xi-\frac{1}{2}\sigma^2\xi^2+\int_{\mathbb{R}}(e^{i\xi\eta}-1-i\xi\eta\mathbf{1}_{[-1,1]}(\eta))\nu_\mathrm{L}(d\eta).
\end{equation}
The L\'evy-Khinchine triplet $(\mu,\sigma,\nu_L)$ uniquely defines the L\'evy process. The value of $\mu$ defines the linear drift of the process, $\sigma$ is the volatility of the diffusion part of the process, and $\nu_\mathrm{L}(\eta)$ is the L\'evy measure, related to the jump part of the process. Under the risk-neutral measure the parameters of the triplet are linked by the equation
\begin{equation}
\mu = r-q-\frac{1}{2}\sigma^2-\int_\mathbb{R}(e^\eta-1-i\eta\mathbf{1}_{[-1,1]}(\eta))\nu_L(d\eta),
\end{equation}
where $r$ is the risk-free interest rate and $q$ is the dividend rate.
In general the characteristic function of a L\'evy process is available in closed form, for example for the Gaussian \citep{Schoutens2003}, NIG \citep{Barndorff1998}, CGMY \citep{Carr2002}, Kou double exponential \citep{kou2002jump}, Merton jump diffusion \citep{merton1976option}, L\'evy alpha stable \citep{Nolan2018}, VG \citep{Madan1990}, Meixner \citep{Schoutens2003}, and mixed tempered stable \citep{Mercuri2016} processes.

Some pricing techniques based on the Fourier transform, e.g.\ FGM and FL, also use the Hilbert transform, which is an integral transform related to the Fourier transform. However, in contrast to the Fourier transform, the function under transformation remains in the same domain, rather than moving between the $x$ and $\xi$ domains. The Hilbert transform of a function in the Fourier domain is defined as
\begin{align}
\label{eq:HilbTrans}
\mathcal{H}\big[\widehat{f}(\xi)\big]&=\,\mathrm{P.V.}\,\frac{1}{\pi}\int^{+\infty}_{-\infty}\frac{\widehat{f}(\xi')}{\xi-\xi'}d\xi'\nonumber\\
&=\lim_{\epsilon\rightarrow0^+}\frac{1}{\pi}\left(\int_{\xi-1/\epsilon}^{\xi-\epsilon} \frac{\widehat{f}(\xi')}{\xi-\xi'} d\xi'+\int_{\xi+\epsilon}^{\xi+1/\epsilon} \frac{\widehat{f}(\xi')}{\xi-\xi'} d\xi'\right),
\end{align}
where $\mathrm{P.V.}$ denotes the Cauchy principal value. Applying the Hilbert transform in the Fourier domain is equivalent to multiplying the function in the $x$ domain by $-i\,\mathrm{sgn}\,x$; see e.g. \cite{King2009,fusai2015}. 

\subsection{Applying barriers with Hilbert transforms}\label{sec:BarrHilb}
The Hilbert transform can be used to obtain the Fourier transform of a function above or below a barrier $b$, i.e.\ $\widehat{f_{b+}}(\xi)=\mathcal{F}_{x\rightarrow\xi}\left[f(x)\mathbf{1}_{\mathbb{R}_+}(x-b)\right]$ and $\widehat{f_{b-}}(\xi)=\mathcal{F}_{x\rightarrow\xi}\left[f(x)\mathbf{1}_{\mathbb{R}_-}(x-b)\right]$, without leaving the Fourier domain via the generalised Plemelj-Sokhotski relations which also employ the shift theorem $\mathcal{F}_{x\rightarrow\xi}[f(x+b)]=\widehat{f}(\xi)e^{-ib\xi}$. These are
\begin{align}
\widehat{f_{b+}}(\xi)&=\frac{1}{2}\big\{\widehat{f}(\xi)+e^{ib\xi}i\mathcal{H}\big[e^{-ib\xi}\widehat{f}(\xi)\big]\big\},\label{eq:PSRelgenpos}\\
\widehat{f_{b-}}(\xi)&=\frac{1}{2}\big\{\widehat{f}(\xi)-e^{ib\xi}i\mathcal{H}\big[e^{-ib\xi}\widehat{f}(\xi)\big]\big\}.\label{eq:PSRelgenneg}
\end{align}
Eqs.~(\ref{eq:PSRelgenpos}) and (\ref{eq:PSRelgenneg}) can be combined to obtain the Fourier transform of the part of a function between two barriers, i.e.\ $\widehat{f_{lu}}(\xi)=\mathcal{F}_{x\rightarrow\xi}\left[f(x)\mathbf{1}_{(l,u)}(x)\right]$,
\begin{equation}
\widehat{f_{lu}}(\xi)=\frac{1}{2}\big\{e^{il\xi}i\mathcal{H}\big[e^{-il\xi}\widehat{f}(\xi)\big]-e^{iu\xi}i\mathcal{H}\big[e^{-iu\xi}\widehat{f}(\xi)\big]\big\}.\label{eq:PSRelgenbarr}
\end{equation}
We may also be required to factorise a function, i.e.\ obtain $\widehat{g_+}(\xi)$ and $\widehat{g_-}(\xi)$ such that $\widehat{g}(\xi)=\widehat{g_+}(\xi)\widehat{g_-}(\xi)$. This is achieved through a log-decompositon, i.e.\ decomposing the logarithm $\widehat{h}(\xi)=\log\widehat{g}(\xi)$ and then exponentiating the results to obtain $\widehat{g_+}(\xi)=\exp\widehat{h_+}(\xi)$ and $\widehat{g_-}(\xi)=\exp\widehat{h_-}(\xi)$.

The Hilbert transform was used by \cite{Feng2008} to price discrete barrier options exploiting the relationship between the price at two successive monitoring dates: being $t_n=n\Delta t,\,n=0,1,\cdots,N,$
\begin{equation}
v(x,t_{n-1}) = \int^u_lv(x',t_n)f_X(x-x',\Delta t)dx',\ n=N,N-1,\cdots,1.
\end{equation}
Here $v(x,t_N)=\phi(x)e^{-\alpha x}$, i.e.\ the payoff of the option, and $f_X(\cdot,\Delta t)$ denotes the transition density of the underlying process with step size $\Delta t$. Using the convolution theorem together with the Hilbert transform, Eqs.~(\ref{eq:PSRelgenpos})--(\ref{eq:PSRelgenbarr}) can be employed to express the relationship between the price at two successive dates as
\begin{equation}
\label{eq:Hilbpricsing}
\widehat{v}(\xi,t_{n-1})=\frac{1}{2}\left\{\Psi(\xi+i\alpha,\Delta t)\widehat{v}(\xi,t_{n})+e^{il\xi}i\mathcal{H}\left[e^{-il\xi}\Psi(\xi+i\alpha,\Delta t)\widehat{v}(\xi,t_{n})\right]\right\}
\end{equation}
for a single-barrier down-and-out option and
\begin{equation}
\label{eq:Hilbpricdoub}
\widehat{v}(\xi,t_{n-1})=\frac{1}{2}\left\{e^{il\xi}i\mathcal{H}\left[e^{-il\xi}\Psi(\xi+i\alpha,\Delta t)\widehat{v}(\xi,t_{n})\right]-e^{iu\xi}i\mathcal{H}\left[e^{-iu\xi}\Psi(\xi+i\alpha,\Delta t)\widehat{v}(\xi,t_{n})\right]\right\}
\end{equation}
for a double-barrier option.
\subsection{Spitzer identities}
If we wish to use Eq.~(\ref{eq:Planch}) to price barrier options, the required characteristic functions are more complicated than the closed-form expressions referred to in Section~\ref{sec:Back_fourhilb}. We need the characteristic function of the distribution of the value of a process $X(t)$ at time $t=T$, conditional on the process remaining inside upper and lower barriers at discrete monitoring dates $t_n$, $n=0,1,\dots,N$. Fortunately, for a single barrier we can use the identities by \cite{spitzer1956combinatorial}, and for double barriers their extension by \cite{Kemperman1963}. These provide the Fourier-$z$ transform of the required PDF: the Fourier transform is applied to the log-price $x$ and the $z$-transform is applied to the discrete monitoring times. The $z$-transform of a discrete function $f(t_n)$ with $n\in\mathbb{N}_0$ is defined as
\begin{equation}
\label{eq:ZFor}
\widetilde{f}(q)=\mathcal{Z}_{n\rightarrow q}[f(t_n)]=\sum_{n=0}^\infty f(t_n)q^n, \quad q\in\mathbb{C}.
\end{equation}
This paper uses the formulation of the Spitzer identities by \cite{Green2010} which was also used by \cite{fusai2015}. The idea is that the required PDFs at subsequent monitoring dates are related as
\begin{equation}
p(x,t_{n-1}) = \int^u_lp(x',t_n)f_X(x-x',\Delta t)dx'.
\end{equation}
where $f_X(x-x',\Delta t)$ is the transition density of the underlying process and $p(x,t_N)=\delta(x)$. By applying the $z$-transform to $p(x,t_N)$ we obtain a Wiener-Hopf style equation which we must solve to obtain the required PDF
\begin{equation}
\widetilde{p}(x,q) - q\int^u_l\widetilde{p}(x',q)f_X(x-x',\Delta t)dx' =\delta(x). \label{eq:GreenWH}
\end{equation}
The Wiener-Hopf method is a large subject and, whilst we briefly outline the technique here to provide a background to the pricing schemes, we refer the interested reader to references such as \cite{Noble1958,Daniele2014} for a detailed treatment of the general method and \cite{Green2010,fusai2015} for a guide to its use in this application. Exploiting the convolution theorem, i.e. $\mathcal{F}_{x\rightarrow\xi}[g(x)*h(x)]=g(\xi)h(\xi)$, Eq.~(\ref{eq:GreenWH}) is transformed into the Fourier domain as
\begin{equation}
\widetilde{\widehat{p}}(\xi,q)(1-q\Psi(\xi,\Delta t))+\widetilde{\widehat{p}}_+(\xi,q)+\widetilde{\widehat{p}}_-(\xi,q) =1,\label{eq:GreenWHFour}
\end{equation}
where $\widetilde{\widehat{p}}_+(\xi,q)$ and $\widetilde{\widehat{p}}_-(\xi,q)$ are (unknown) auxiliary functions defined to extend Eq.~(\ref{eq:GreenWH}) along the entire $x$-axis. To solve this for a single (lower) barrier we have $\widetilde{\widehat{p}}_+(\xi,q)=0$. We first factorise $\Phi(\xi,q)=1-q\Psi(\xi,\Delta t)=\Phi_+(\xi,q)\Phi_-(\xi,q)$ and divide both sides by $\Phi_-(\xi,q)$ so that the wanted ``+" function $\widetilde{\widehat{p}}(\xi,q)$ is multiplied with the ``+" function $\Phi_+(\xi,q)$ only. We then discard $\widetilde{\widehat{p}}_-(\xi,q)/\Phi_-(\xi,q)$ which is purely negative and decompose $1/\Phi_-(\xi,q)$ around $l$ to obtain the positive function $P_{l+}(\xi,q)=[1/\Phi_-(\xi,q)]_{l+}$. The required PDF function subject to monitoring at a lower barrier is
\begin{equation}
\widetilde{\widehat{p}}(\xi,q)=\frac{P_{l+}(\xi,q)}{\Phi_+(\xi,q)},
\end{equation}
which corresponds to \cite[Eq.~(14)]{fusai2015}.\footnote{With reference to the notation in \cite{fusai2015}, $[1/\Phi_-(\xi,q)]_{l+}=[e^{-il\xi}/\Phi_-(\xi,q)]_{+}$, due to the shift theorem.}
For double barrier options, the procedure is more complicated as both of the auxiliary functions $\widetilde{\widehat{p}}_+(\xi,q)$ and $\widetilde{\widehat{p}}_-(\xi,q)$ are non zero. We calculate the Fourier-$z$ transform of the wanted PDF by subtracting the unwanted parts above and below the barriers, i.e.
\begin{equation}
\widetilde{\widehat{p}}(\xi,q)=\frac{1-J_{l-}(\xi,q)-J_{u+}(\xi,q)}{\Phi(\xi,q)}
\end{equation}
which corresponds to \cite[Eq.~(16)]{fusai2015}.
However, the calculation of $J_{u+}(\xi,q)$ requires the knowledge of $J_{l-}(\xi,q)$ and vice versa. In fact they are linked by the following coupled equations which so far have only been solved by an iterative method:
\begin{align}
\frac{J_{l-}(\xi,q)}{\Phi_-(\xi,q)} &= \left[\frac{1-J_{u+}(\xi,q)}{\Phi_-(\xi,q)}\right]_{l-},\label{eq:Jmin} \\
\frac{J_{u+}(\xi,q)}{\Phi_+(\xi,q)} &= \left[\frac{1-J_{l-}(\xi,q)}{\Phi_+(\xi,q)}\right]_{u+}.\label{eq:Jpos}
\end{align}

\subsection{Numerical methods}\label{sec:Back_num}
The methods in the previous section are described analytically. However, as they involve some expressions which cannot be solved in closed form, their implementation requires the use of numerical approximation techniques which we discuss in the following.

\subsubsection{Discrete Fourier transform and spectral filtering}\label{sec:Nummeth_DFT}
The forward and reverse Fourier transforms in Eqs.~(\ref{eq:FwdFourier}) and (\ref{eq:RevFourier}) are integrals over an infinite domain and in order to compute them numerically one needs to approximate them with a discrete Fourier transform (DFT). Rather than being defined over an infinite and continuous range of $x$ and $\xi$ values, the DFT is defined on grids of size $M$ in the $x$ and $\xi$ domains. For our scheme both the $x$ and $\xi$ grids are centred around zero and are defined based on the maximum value in the $x$ domain $x_{\max}$. The step size is $\Delta x=2x_{\max}/M$ and the $x$ domain grid is defined as
\begin{equation}
\label{eq:xGrid}
x_j=j\Delta x,\quad j=-\frac{M}{2},-\frac{M}{2}+1,\dots,\frac{M}{2}-1.
\end{equation}
The points in the $\xi$ domain are then calculated according to the Nyquist relation by obtaining the step size $\Delta\xi=\pi/x_{\max}$ and range $\xi_{\max}=\pi/\Delta x$ to give the $\xi$ domain grid as
\begin{equation}
\label{eq:xiGrid}
\xi_k=k\Delta\xi, \quad k=-\frac{M}{2},-\frac{M}{2}+1,\dots,\frac{M}{2}-1.
\end{equation}
The discrete Fourier transform is then
\begin{align}
\widehat{f}_{M,\Delta x}(\xi_k)&=\Delta x \sum^{M/2-1}_{j=-M/2}f\left(x_j\right)e^{ix_j\xi_k}, \label{eq:DFT1}\\
f_{M,\Delta\xi }(x_j)&=\frac{\Delta\xi}{2\pi}\sum^{M/2-1}_{k=-M/2}\widehat{f}\left(\xi_k\right)e^{-ix_j\xi_k}.\label{eq:revDFT1}
\end{align}
In practice, we perform this calculation using the built-in MATLAB FFT function based on the FFTW library by \cite{frigo1998fftw}.

\begin{figure}
\begin{center}
\includegraphics[width=\textwidth]{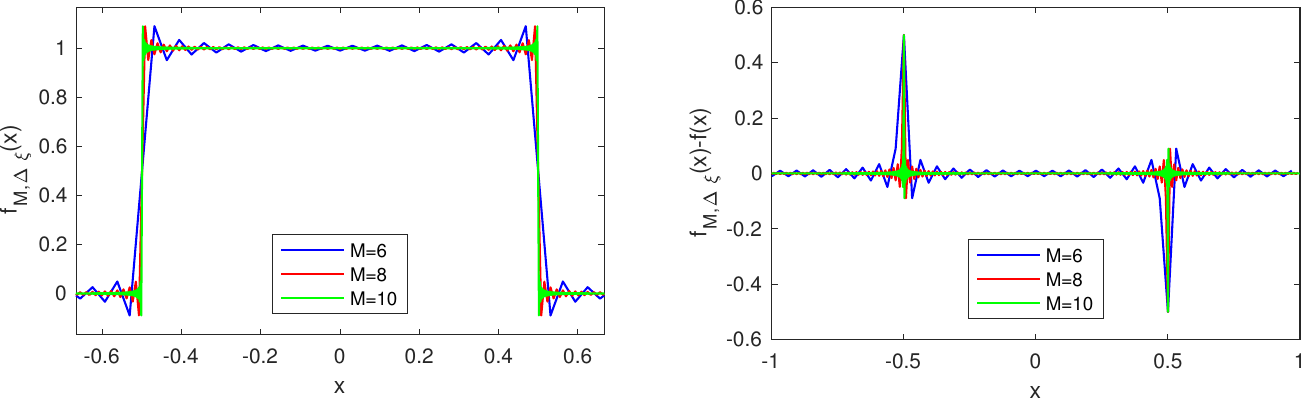}
\caption{Illustration of the effect of the Gibbs phenomenon on a rectangular pulse recovered applying the inverse FFT with grid size $M$ to $\mathrm{sinc}(\xi/2\pi)$. The approximated function is shown on the left and the error with respect to an analytical rectangular pulse on the right. On increasing $M$, the peak error at the discontinuity remains the same, the error away from the discontinuity reduces and the frequency of the oscillations increases.}
\label{fig:pulsegibbs}
\end{center}
\end{figure}

It can be seen in Eqs.~(\ref{eq:DFT1}) and (\ref{eq:revDFT1}) that the range over which we calculate the Fourier transform is truncated, so we must consider the effect of the Gibbs phenomenon on the error convergence. This describes the way that the finite sum in Eq.~(\ref{eq:revDFT1}) converges to the analytical function $f(x)$ corresponding to an infinite sum. \cite{hewitt1979gibbs} provided a comprehensive guide to this effect which was first observed by \cite{wilbraham1848certain} and later described by \cite{gibbs1898fourier,gibbs1899fourier}. 
An example of this can be seen in Figure~\ref{fig:pulsegibbs} which shows how $f_{M,\Delta\xi}(x)$ for a rectangular pulse varies as the value of $M$ increases. The error peaks at the discontinuity $f(x_\mathrm{d})$ and oscillates away from it, with the amplitude decreasing as a function of distance from the discontinuity.
The value of the recovered function at the discontinuity $f_{M,\Delta\xi}(x_\mathrm{d})$, where $|x_\mathrm{d}|=0.5$, will be the mean of the values immediately before and after the discontinuity, i.e.\ $f_{M,\Delta\xi}(x_\mathrm{d})=\frac{1}{2}[f(x_\mathrm{d}^+)+f(x_\mathrm{d}^-)]$, and thus stays the same even as the value of $M$ increases. In contrast, it can be observed from Figure~\ref{fig:pulsegibbs} that the oscillations increase in frequency and decrease in amplitude as the value of $M$ increases.

However, the most important aspect of the Gibbs phenomenon for the work in this paper is the way that the shape of the function in the Fourier domain is related to its shape in the state space. From the ``integration by parts coefficient bound'' described by \cite{Boyd2001} \citep[see also][]{ruijter2015application}, if the function is smooth up to and including its $(k-2)^{\mathrm{th}}$ derivative, and its $k^{\mathrm{th}}$ derivative is integrable, then the Fourier coefficients decrease as $O(1/\xi^k)$.

\begin{figure}
\begin{center}
\includegraphics[width=\textwidth]{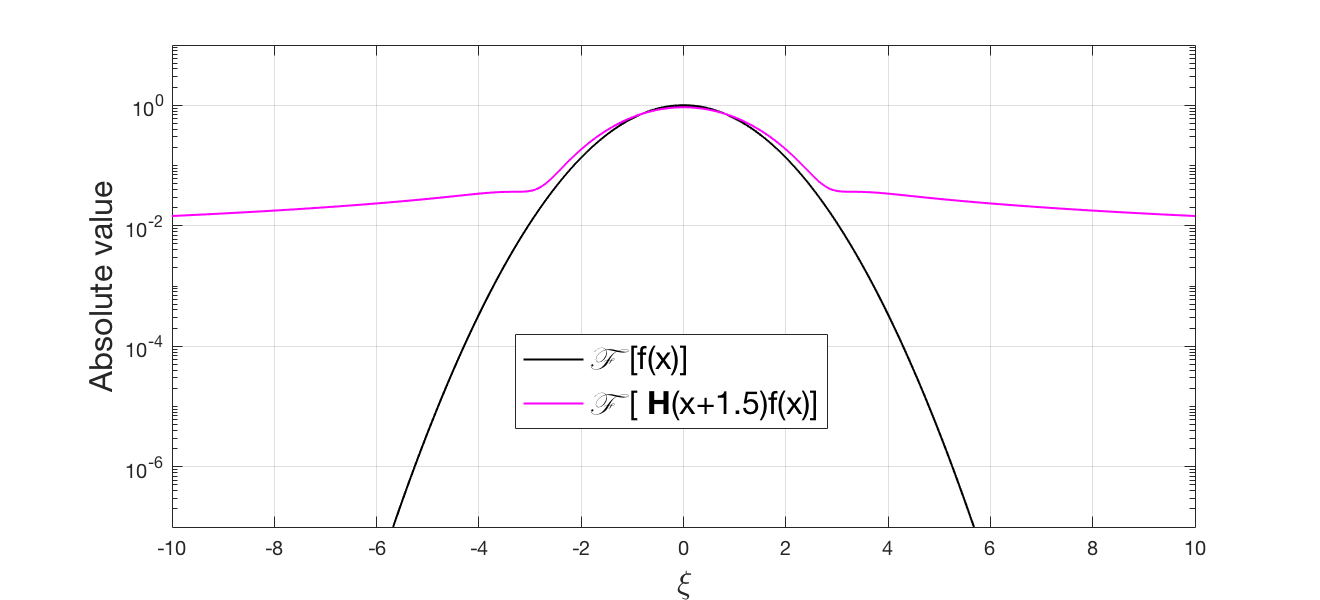}
\caption{Illustration of the effect of the Gibbs phenomenon on a the Fourier transform of a standard normal distribution. It can be seen that introducing a step into a function has the effect of changing the decay of the Fourier transform from exponential to $O(1/\xi)$; here \textbf{H}$(x)$ denotes the Heaviside step function.
}
\label{fig:Gibbsplot}
\end{center}
\end{figure}

This is illustrated in Figure~\ref{fig:Gibbsplot} which shows the Fourier transform of a standard normal distribution and the Fourier transform of a standard normal distribution multiplied by a step at $x=-1.5$.


Many proposed solutions to the Gibbs phenomenon are too computationally heavy to be useful for our application, such as adaptive filtering and mollifiers suggested by \cite{tadmor2005adaptive} and \cite{tadmor2007filters}. Moreover, as explained in Section~\ref{sec:Errperf}, our error analysis is particularly concerned with the shape of the function in the Fourier domain. Therefore we adopt the approach of \cite{ruijter2015application}, using simple spectral filtering techniques which are applied by a pointwise multiplication in the Fourier domain and therefore shape the function directly whilst adding very little computational load.

In the papers by \cite{vandeven1991family} and \cite{gottlieb1997gibbs}, a filter of order $p$ is defined as a function $\sigma(\eta)$ supported on $\eta\in[-1,1]$ with the following properties:
\begin{align}
\label{eq:filtDef}
& \text{a) }\sigma(0)=1\text{,     } \sigma^{(l)}(0)=0 \nonumber\\
& \text{b) }\sigma(\eta)=0 \text{ for }|\eta|=1\nonumber\\
& \text{c) }\sigma(\eta)\in C^{p-1}.
\end{align}
The scaled variable $\eta$ is related to $\xi$ in our application as $\eta=\xi/\xi_{\max}$. In this paper we investigate the use of two filters. The exponential filter, described by \cite{gottlieb1997gibbs} has the form
\begin{equation}
\label{eq:expFilt}
\sigma(\eta)=e^{-\vartheta\eta^p},
\end{equation}
where $p$ is even and positive. This does not strictly meet criterion b in Eq.~(\ref{eq:filtDef}) as it does not go exactly to zero when $|\eta|=1$. However, if we select $\vartheta<\varepsilon\log 10$, where $10^{-\varepsilon}$ is machine precision, then the filter coefficients are within computational accuracy of the requirements. An advantage of the exponential filter is that it has a simple form and the order of the filter is equal to the parameter $p$ which is directly input to the filter equation.

\begin{figure}
\begin{center}
\includegraphics[width=\textwidth]{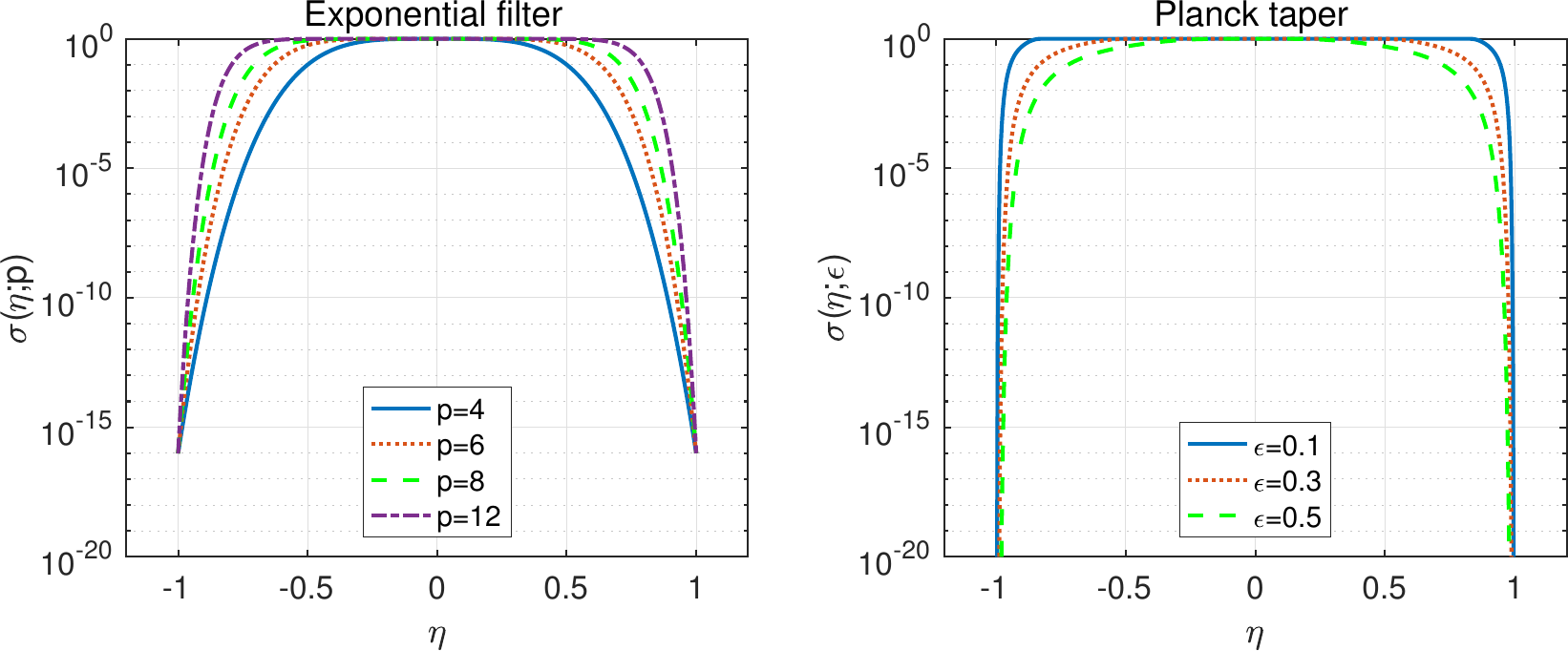}
\caption{Shape of the exponential filter (left) and Planck taper (right) with different parameter values.}
\label{fig:expptfilt}
\end{center}
\end{figure}

The other filter we study here is the Planck taper \citep{mckechan2010tapering}, which is defined piecewise as
\begin{equation}
\sigma(\eta)=
\begin{cases}
0, &\eta\leq \eta_1,\ \ \qquad \eta_1=-1, \\
\frac{1}{e^{\;z(\eta)}+1},\ z(\eta)=\frac{\eta_2-\eta_1}{\eta-\eta_1}+\frac{\eta_2-\eta_1}{\eta-\eta_2}, & \eta_1<\eta<\eta_2,\ \, \eta_2=\epsilon-1,\\
1, & \eta_2 \leq\eta\leq\eta_3,\ \,\eta_3=1-\epsilon,\\
\frac{1}{e^{\;z(\eta)}+1},\ z(\eta)=\frac{\eta_3-\eta_4}{\eta-\eta_3}+\frac{\eta_3-\eta_4}{\eta-\eta_4}, &\eta_3<\eta<\eta_4,\ \, \eta_4=1,\\
0, &\eta\geq\eta_4.
\end{cases}
\end{equation}
The value of $\epsilon$ gives the proportion of the range of $\eta$ which is used for the slope regions. Outside these regions, it is completely flat with a value of $1$. This contrasts with the exponential filter which introduces some, albeit often very minor, distortion for any value of $\eta\neq0$. In addition the Planck taper has the notable property that for all values of $\epsilon>0$, $\sigma(\eta,\epsilon)\in C^\infty$ and therefore the order of the Planck taper is $\infty$. However, it is clear that different values of $\epsilon$ give a different filter shape, so the order of a filter alone cannot be taken as a predictor of performance. Examples of the two filters are shown in Figure~\ref{fig:expptfilt}.
\FloatBarrier

\subsubsection{Hilbert transform}\label{Sec:Back_transf_Hilb}
The calculation of the Hilbert transform of a function $\widehat{f}(\xi)$ can be realised with an inverse/forward Fourier transform pair and multiplication by the signum function,
\begin{equation}
\mathcal{H}\big[\widehat{f}(\xi)\big]=-i\,\mathcal{F}_{x\rightarrow\xi}\big[\mathrm{sgn}(x)\mathcal{F}^{-1}_{\xi\rightarrow x}\widehat{f}(\xi)\big].
\end{equation}
However, this gives an error convergence which is polynomially decreasing with the number of grid points $M$. In order to obtain exponential error convergence, \cite{Feng2008} and \cite{fusai2015} have implemented the Hilbert transform using the sinc expansion techniques comprehensively studied by \cite{Stenger1993,Stenger2011}. Stenger showed that, given a function $\widehat{f}(\xi)$ which is analytic in the whole plane including the real axis, the function and its Hilbert transform can be expressed as
\begin{align}
\widehat{f}(\xi)&=\sum^{+\infty}_{k=-\infty}\widehat{f}(k\Delta\xi)\frac{\sin(\pi(\xi-k\Delta\xi)/\Delta\xi)}{\pi(\xi-k\Delta\xi)/\Delta\xi},\label{eq:SincApprox}\\
\mathcal{H}\big[\widehat{f}(\xi)\big]&=\sum^{+\infty}_{k=-\infty}\widehat{f}(k\Delta\xi)\frac{1-\cos(\pi(\xi-k\Delta\xi)/\Delta\xi)}{\pi(\xi-k\Delta\xi)/\Delta\xi}, \label{eq:HilbSincApprox}
\end{align}
where $\Delta\xi$ is the grid step size in the Fourier domain. \cite{Stenger1993} also showed that, when the function $\widehat{f}(\xi)$ is analytic in a strip of the complex plane including the real axis, the expressions in Eqs.~(\ref{eq:SincApprox}) and (\ref{eq:HilbSincApprox}) are approximations whose error decays exponentially as $\Delta\xi$ decreases. In addition to discretisation, the infinite sum in Eq.~(\ref{eq:HilbSincApprox}) must also be truncated to the grid size $M$, so that the discrete approximation of the Hilbert transform becomes
\begin{equation}
\label{eq:HilbSincApproxTrunc}
\mathcal{H}\big[\widehat{f}(\xi)\big]\approx\sum^{+M/2}_{k=-M/2}\widehat{f}(k\Delta\xi)\frac{1-\cos(\pi(\xi-k\Delta\xi)/\Delta\xi)}{\pi(\xi-k\Delta\xi)/\Delta\xi}.
\end{equation}
\cite{Feng2008,Feng2009} showed that if $\widehat{f}(\xi)$ decays at least exponentially as $\xi\rightarrow\infty$, i.e.\ $\widehat{f}(\xi)\leq \kappa \exp(-c |\xi|^{\nu})$, then the error in the Hilbert transform and the Plemelj-Sokhotski relations caused by truncating the infinite sum in Eq.~(\ref{eq:HilbSincApprox}) is also exponentially bounded. Furthermore Feng and Linetsky showed that if $\widehat{f}(\xi)$ is polynomially bounded as $\xi\rightarrow\infty$, i.e.\ $\widehat{f}(\xi)\leq c |\xi|^{\nu}$, then the error caused by truncating the series is no longer exponentially bounded \citep{Feng2008,Feng2009}.

\subsubsection{Pricing method: single-barrier options with the Spitzer identity}\label{sec:back_num_sing}
Two of the pricing methods that we modify in order to reduce the errors from the discrete Hilbert transform were devised and explained in depth by \cite{fusai2015}. The first method which we examine is the pricing procedure for single-barrier options. Without loss of generality we consider only the down-and-out case; the modifications that we propose are equally applicable to other types of single-barrier options. This method is briefly described here in order to provide a backdrop to the changes that were made to improve convergence.

\begin{enumerate}
\item Set the number of dates to $N-2$ so that the characteristic function acts as a smoothing function for the first and last dates in the scheme.
\item Compute the characteristic function $\Psi(\xi+i\alpha,\Delta t)$, where $\alpha$ is the damping factor used in Section~\ref{sec:Back_fourhilb}.
\item Use the Plemelj-Sokhotski relations with the sinc method to factorise
\begin{equation}
\label{eq:Phifactsing}
\Phi(\xi,q):=1-q\Psi(\xi+i\alpha,\Delta t) =\Phi_+(\xi,q)\Phi_-(\xi,q)
\end{equation}
with $q$ selected according to the criteria specified by \cite{Abate1992} for the inverse $z$-transform.
\item Decompose
\begin{equation}
\label{eq:Pdecompsing}
P(\xi,q):=\Psi(\xi+i\alpha,\Delta t)/\Phi_-(\xi,q) = P_{l+}(\xi,q)+P_{l-}(\xi,q)
\end{equation}
and calculate
\begin{equation}
\label{eq:FinalfpCalc1}
F(\xi,q):=\widehat{\phi}\,^*(\xi)\Psi(\xi+i\alpha,\Delta t)\frac{P_{l+}(\xi,q)}{\Phi_+(\xi,q)}.
\end{equation}
\item Calculate the price
\begin{equation}
\label{eq:FinalpriceCalc}
v(0,N):=e^{-rT}\mathcal{F}^{-1}_{\xi\rightarrow x=0}\mathcal{Z}^{-1}_{q\rightarrow n= N-2}[F(\xi,q)].
\end{equation}
\end{enumerate}
The Spitzer identities give the $z$-transform of the characteristic function, so the inverse $z$-transform $\mathcal{Z}^{-1}_{q\rightarrow n}[\widetilde{f}(q)]$ must be applied. The method used was devised by \cite{Abate1992_2,Abate1992} and approximates the inverse $z$-transform by
\begin{equation}
f(t_n)\approx\frac{1}{2^{m_\mathrm{E}}n\rho^n} \sum^{m_\mathrm{E}}_{j=0}\binom{m_\mathrm{E}}{j}b_{n_\mathrm{E}+j},
\end{equation}
where
\begin{equation}
\label{eq:AWInvZpartial}
b_k=\frac{1}{2}\widetilde{f}(\rho)+\sum^k_{j=1}(-1)^j\operatorname{Re}\widetilde{f}\Big(\rho e^{\frac{\pi j i}{n}}\Big).
\end{equation}

The parameters $n_\mathrm{E}$ and $m_\mathrm{E}$ are chosen to be large enough to attain sufficient accuracy and small enough such that $n_\mathrm{E}+m_\mathrm{E}\ll n$. Tests suggest that a choice of $n_\mathrm{E}=12$ and $m_\mathrm{E}=20$ provides good accuracy. This gives $n_\mathrm{E}+m_\mathrm{E}=32$ which is much smaller than the number of dates specified in most option contracts.
The parameter $\rho$ controls the accuracy of the inverse $z$-transform; in order to have an accuracy of $10^{-2\gamma}$, one must set $\rho=10^{-\gamma/n}$ \citep{Abate1992}. This can result in very small values of $\rho$ and so it has been found in practice that the best achievable performance is of the order of $10^{-12}$ with $\gamma=6$. However, this is more than sufficiently low for practical purposes and to show whether exponential convergence is achieved.

\cite{fusai2015} showed that this method could achieve exponential convergence with a wide range of L\'evy processes. However, with the VG process this method only achieved polynomial convergence. This is consistent with the error behaviour of the discrete Hilbert transform with the VG process, as explained in Section~\ref{Sec:Back_transf_Hilb} above. Section~\ref{sec:Errperf} explains in more detail how the error convergence is bounded when this process is used.

In order to improve the result,
we multiplied the characteristic function by a spectral filter $\sigma(\eta)$ so that the input to both the factorisation and decomposition steps decay exponentially. The expressions in Eqs.~(\ref{eq:Phifactsing}) and (\ref{eq:Pdecompsing}) are replaced by
\begin{align}
\Phi(\xi,q) & := 1-q\Psi(\xi+i\alpha,\Delta t)\sigma\left(\xi/\xi_{\max}\right)=\Phi_+(\xi,q)\Phi_-(\xi,q),\label{eq:Phifactsingfilt} \\
P(\xi,q) & := \frac{\Psi(\xi+i\alpha,\Delta t)\sigma\left(\xi/\xi_{\max}\right)}{\Phi_-(\xi,q)} = P_{l+}(\xi,q)+P_{l-}(\xi,q). \label{eq:Pdecompsingfilt}
\end{align}

\subsubsection{Pricing method: double-barrier options with the Spitzer identity}\label{sec:back_num_doub}
The second method from \cite{fusai2015} which we examine in this paper is the pricing procedure for double-barrier options. This is very similar to the method for the single-barrier options described in Section~\ref{sec:back_num_sing}, in that it uses Wiener-Hopf factorisation and decomposition to compute the appropriate Spitzer identitiy. However, the major difference in this case is that the equations cannot be solved directly and so require the use of a fixed-point algorithm. The steps in the pricing procedure for double-barrier options are the same as the procedure described for the single-barrier down-and-out option described in Section~\ref{sec:back_num_sing} with the exception of Step 4 which is now replaced by the following fixed-point algorithm:
\begin{enumerate}
\item[4. (a)] Set $J_{u+}(\xi,q)=J_{l-}(\xi,q)=0$.
\item[\phantom{4. }(b)] Decompose
\begin{align}
\label{eq:Pdecomp}
& P(\xi,q):=\frac{\Psi(\xi+i\alpha,\Delta t)-J_{u+}(\xi,q)}{\Phi_-(\xi,q)} = P_{l+}(\xi,q)+P_{l-}(\xi,q)
\end{align}
and set $J_{l-}(\xi,q):=P_{l-}(\xi,q)\Phi_-(\xi,q)$.
\item[\phantom{4. }(c)] Decompose
\begin{align}
\label{eq:Qdecomp}
& Q(\xi,q):=\frac{\Psi(\xi+i\alpha,\Delta t)-J_{l-}(\xi,q)}{\Phi_+(\xi,q)} = Q_{u+}(\xi,q)+Q_{u-}(\xi,q)
\end{align}
and set $J_{u+}(\xi,q):=Q_{u+}(\xi,q)\Phi_+(\xi,q)$.
\item[\phantom{4. }(d)] Calculate
\begin{align}
\label{eq:FinalfpCalc2}
F(\xi,q):=\widehat{\phi}\,^*(\xi)\frac{\Psi(\xi+i\alpha,\Delta t)}{\Phi(\xi,q)}\left[\Psi(\xi+i\alpha,q)-J_{l-}(\xi,q)-J_{u+}(\xi,q)\right].
\end{align}
\item[\phantom{4. }(e)] If the difference between the new and the old value of $F(\xi,q)$ is less than a predefined tolerance or the number of iterations is greater than a certain value, e.g.~5, then calculate the price using Eq.~(\ref{eq:FinalpriceCalc}), otherwise return to step (b).
\end{enumerate}

Unlike the direct method for single-barrier options described in Section~\ref{sec:back_num_sing}, this iterative method is limited to polynomial error convergence for all processes. In Section~\ref{sec:Errperf} we show that this is due to the Gibbs phenomenon.
In order to improve the error convergence we placed a filter $\sigma(\eta)$ on the input to each decomposition step in the fixed-point algorithm. The calculations for $P(\xi,q)$ and $Q(\xi,q)$ in Eqs.~(\ref{eq:Pdecomp}) and (\ref{eq:Qdecomp}) are replaced by
\begin{align}
P(\xi,q)&:=\sigma\left(\frac{\xi}{\xi_{\max}}\right)\left[\frac{\Psi(\xi+i\alpha,\Delta t)-J_{u+}(\xi,q)}{\Phi_-(\xi,q)} \right] ,\\
Q(\xi,q)&:=\sigma\left(\frac{\xi}{\xi_{\max}}\right)\left[\frac{\Psi(\xi+i\alpha,\Delta t)-J_{l-}(\xi,q)}{\Phi_+(\xi,q)} \right].
\end{align}

It must also be noted that this change is only designed to provide significant improvements to the double-barrier method with exponentially decaying characteristic functions. In the case of a polynomially decaying characteristic function such as that of the VG process, this method will also be subject to the same limitations on accuracy as described in Section~\ref{sec:Errperf_Sing} for single-barrier options. Therefore, if we wish to use this scheme with the VG process, we must also apply filtering to the factorisation step as shown in Eq.~(\ref{eq:Phifactsingfilt}).
\subsubsection{Pricing method: Feng and Linetsky}\label{sec:back_num_fl}
The third pricing method that we examine in order to illustrate the improvements obtained by the addition of spectral filtering to the sinc-based Hilbert transform is the recursive one published by \cite{Feng2008} and explained in Section~\ref{sec:BarrHilb}.
%
%
In general, the FL method achieves excellent results for both single and double-barrier options \citep{Feng2008,fusai2015}; the error converges exponentially with grid size and reaches machine accuracy for fairly small grid sizes. However, with respect to the FGM model, it has the disadvantage that the computational time increases linearly with the number of monitoring dates.

Similarly to the FGM method for single-barrier options, exponential error convergence is achieved only for processes where the characteristic function reduces exponentially as $|\xi|\rightarrow\infty$. Therefore, poor error convergence is achieved for the VG process which has a characteristic function which only reduces polynomially as $|\xi|\rightarrow\infty$. \cite{Feng2008} explained this in some detail, showing how this is linked to the truncation error of the discrete Hilbert transform. In order to improve the results, we altered the FL method by placing a filter on the input to the Hilbert transform to ensure it decays exponentially. We replaced Eqs.~(\ref{eq:Hilbpricsing}) and (\ref{eq:Hilbpricdoub}) by
\begin{align}
\widehat{v}(\xi,t_{n-1})=&\frac{1}{2}\left\{\sigma\left(\frac{\xi}{\xi_{\max}}\right)\Psi(\xi+i\alpha,\Delta t)\widehat{v}(\xi,t_{n})\right.\nonumber\\&\left.+e^{il\xi}i\mathcal{H}\left[e^{-il\xi}\sigma\left(\frac{\xi}{\xi_{\max}}\right)\Psi(\xi+i\alpha,\Delta t)\widehat{v}(\xi,t_{n})\right]\right\},\label{eq:Hilbpricsingfilt}\\
\widehat{v}(\xi,t_{n-1})=&\frac{1}{2}\left\{e^{il\xi}i\mathcal{H}\left[e^{-il\xi}\sigma\left(\frac{\xi}{\xi_{\max}}\right)\Psi(\xi+i\alpha,\Delta t)\widehat{v}(\xi,t_{n})\right]\right.\nonumber\\&\left.-e^{iu\xi}i\mathcal{H}\left[e^{-iu\xi}\sigma\left(\frac{\xi}{\xi_{\max}}\right)\Psi(\xi+i\alpha,\Delta t)\widehat{v}(\xi,t_{n})\right]\right\}.\label{eq:Hilbpricdoubfilt}
\end{align}

\section{Error convergence of the pricing procedure}\label{sec:Errperf}
In this section we present new bounds for the error convergence of the different calculations that make up the original pricing procedures without spectral filters and show bounds for the individual steps. In doing this, the effect of each step in the procedure on the shape of the output function in the Fourier domain is examined, as this
largely determines the error convergence of the successive steps. In the FGM and FL pricing methods, the computation of the characteristic function is done directly in the Fourier domain so there are no numerical errors associated with this calculation. All the L\'evy processes that we are considering have characteristic functions that decay exponentially as $|\xi|\rightarrow\infty$, with the exception of the VG process where the characteristic function decays polynomially and is bounded by $|\xi|^{-2\Delta t/\nu}$. The damping factor $\alpha$ is omitted from the calculations to make the notation more concise. This is appropriate as the value of $i\alpha$ becomes insignificant as $|\xi|\rightarrow\infty$. In the error calculations below $c_n$ where $n=1,2,,3...$ are included as positive constants.

\subsection{Pricing single-barrier options with the VG process using the Spitzer identity}\label{sec:Errperf_Sing}
Following the calculation of the characteristic function, the next step in the pricing procedure is the factorisation of $\Phi(\xi,q)=[1-q\Psi(\xi,\Delta t)]$, which means that we need to apply the discrete Hilbert transform to $\log\Phi(\xi,q)=\log[1-q\Psi(\xi,\Delta t)]$. With the exception of the VG process, as $|\xi|\rightarrow\infty$, $q\Psi(\xi,\Delta t)\sim qe^{-\Delta t\xi^2}$ which quickly becomes very small. Thus we can say that as $|\xi|\rightarrow\infty$, $|\log[1-q\Psi(\xi,\Delta t)]|<ce^{-\kappa\Delta t\xi^2}$ with $c,\kappa$ positive constants. Therefore, from the error bounds for the sinc-based Hilbert transform proved by \cite{Stenger1993} and \cite{Feng2008}, the output of the decomposition of $\log\Phi(\xi,q)$ has exponential error convergence for exponentially decaying characteristic functions.

For the VG process the characteristic function is
\begin{equation}
\Psi(\xi,t) =\left(1-i\nu\theta\xi+\frac{1}{2}\nu^2\xi^2\sigma^2\right)^{-t/\nu}.
\end{equation}
When $\xi\to\infty$ then $\Psi(\xi,\Delta t)$ is dominated by $\xi^{-2\Delta t/\nu}$, so $|\log[1-q\Psi(\xi,\Delta t)]|<c\xi^{-2\Delta t/\nu}$. Therefore, we can bound the truncation error from the decomposition of $\log[1-q\Psi(\xi,\Delta t)]$. \cite{Feng2008} showed that the truncation error from applying the sinc-based Hilbert transform to a function which decays as $c|\xi|^{-2\Delta t/\nu}$ is bounded by $\frac{2c\nu}{2\Delta t-\nu}(M\Delta\xi)^{-(2\Delta t/\nu-1)}$, where there is a constraint on the process parameters of $\Delta t>\nu/2$. We show that if we take into account the form of the discrete Hilbert transform and the similarity between the positive and negative tails of the characteristic function, a tighter bound can be defined and the constraints on the parameters can be relaxed. Defining $f_{\Delta\xi}(\xi)$ as the output of the infinite sum from Eq.~(\ref{eq:HilbSincApprox}) and $f_{\Delta\xi,M}(\xi)$ as the output of the truncated sum from Eq.~(\ref{eq:HilbSincApproxTrunc}),
\begin{align}
|f_{\Delta\xi}(\xi)-f_{\Delta\xi,M}(\xi)| <& c_1\Delta\xi\left|\sum_{k>M/2}\frac{(k\Delta\xi)^{-2\Delta t/\nu}\left(1-\cos\left(\pi\frac{\xi-k\Delta\xi}{\Delta\xi}\right)\right)}{\xi-k\Delta\xi}\right.\nonumber\\
& \left. + \sum_{k<-M/2}\frac{(k\Delta\xi)^{-2\Delta t/\nu}\left(1-\cos\left(\pi\frac{\xi-k\Delta\xi}{\Delta\xi}\right)\right)}{\xi-k\Delta\xi}\right|\nonumber\\
<&c_1\Delta\xi\sum_{k>M/2}(k\Delta\xi)^{-2\Delta t/\nu}\left|\frac{1-\cos\left(\pi\frac{\xi-k\Delta\xi}{\Delta\xi}\right)}{\xi-k\Delta\xi}+\frac{1-\cos\left(\pi\frac{\xi+k\Delta\xi}{\Delta\xi}\right)}{\xi+k\Delta\xi} \right|\nonumber\\
<&c_2\Delta\xi\sum_{k>M/2}(k\Delta\xi)^{-2\Delta t/\nu}\left|\frac{1}{\xi-k\Delta\xi}+\frac{1}{\xi+k\Delta\xi} \right|\nonumber\\
<&c_3\Delta\xi\sum_{k>M/2}\frac{(k\Delta\xi)^{-2\Delta t/\nu}}{|\xi^2-(k\Delta\xi)^2|}<c_4\Delta\xi\sum_{k>M/2}\frac{(k\Delta\xi)^{-2\Delta t/\nu}}{(k\Delta\xi)^2}\nonumber\\
<&c_5\int^{+\infty}_{M\Delta\xi/2}\xi_k^{-\left(\frac{2\Delta t}{\nu}+2\right)}d\xi_k
<c_6(M\Delta\xi)^{-\left(\frac{2\Delta t}{\nu}+1\right)}.\label{eq:vgfactbound1}
\end{align}
The two cosines are equal because the difference of their arguments is $2k\pi$.
For the integral to converge we must have $2\Delta t/\nu+2>1$, which is the case for all possible process parameters.
When the output of this decomposition is exponentiated to obtain the results of the factorisation, the error will be bounded as
\begin{align}
\left|\frac{e^{f_{\Delta\xi}(\xi)}-e^{f_{\Delta\xi,M}(\xi)}}{e^{f_{\Delta\xi}(\xi)}}\right| & < c_7\left[1-e^{c_6(M\Delta\xi)^{-\left(\frac{2\Delta t}{\nu}+1\right)}}\right].
\end{align}
For large $M$ this converges as $O\big(M^{-\left(2\Delta t/\nu+1\right)}\big)$; thus the error convergence of the factorisation is polynomial. The expression in Eq.~(\ref{eq:vgfactbound1}) gives the error at fixed values of $\xi$, i.e.\ the chosen grid points; therefore $\xi$ can be absorbed into $c_3$ and $c_4$. Moreover, in the final price calculation the Spitzer identities are multiplied by the payoff and the characteristic function which both decay as $|\xi|$ increases and therefore the errors close to $\xi=0$ will have the largest influence on the final error of the solution. However, as $M$ increases, our range of $\xi$ increases and so we should consider the effect of errors at large values of $\xi$ on the error of the next step in the calculation. 
As explained below, we multiply the input to the subsequent Hilbert transform by the characteristic function and if the number of grid points increases from $M^*$ to $M$, where $M^*\ll M$, then the additional error from these points is bounded by
\begin{equation}
c_8\sum^{M}_{j=M^*}\frac{j\Delta \xi M^{-\left(4\Delta t/\nu+1\right)}}{\xi-j\Delta \xi}<c_9\sum^{M}_{j=M^*}M^{-\left(4\Delta t/\nu+1\right)}<c_{10} M^{-\left(4\Delta t/\nu\right)}.
\end{equation}
Depending on the parameters of the VG process, this decreases more or less rapidly than the original bound, and if we were to select our parameters according to the requirement in \cite{Feng2008} of $2\Delta t/\nu>1$ then $O\big(M^{-\left(2\Delta t/\nu+1\right)}\big)$ will dominate. However, regardless of parameter selection, the error converges polynomially with $M$.

\begin{figure}
\begin{center}
\includegraphics[width=\textwidth]{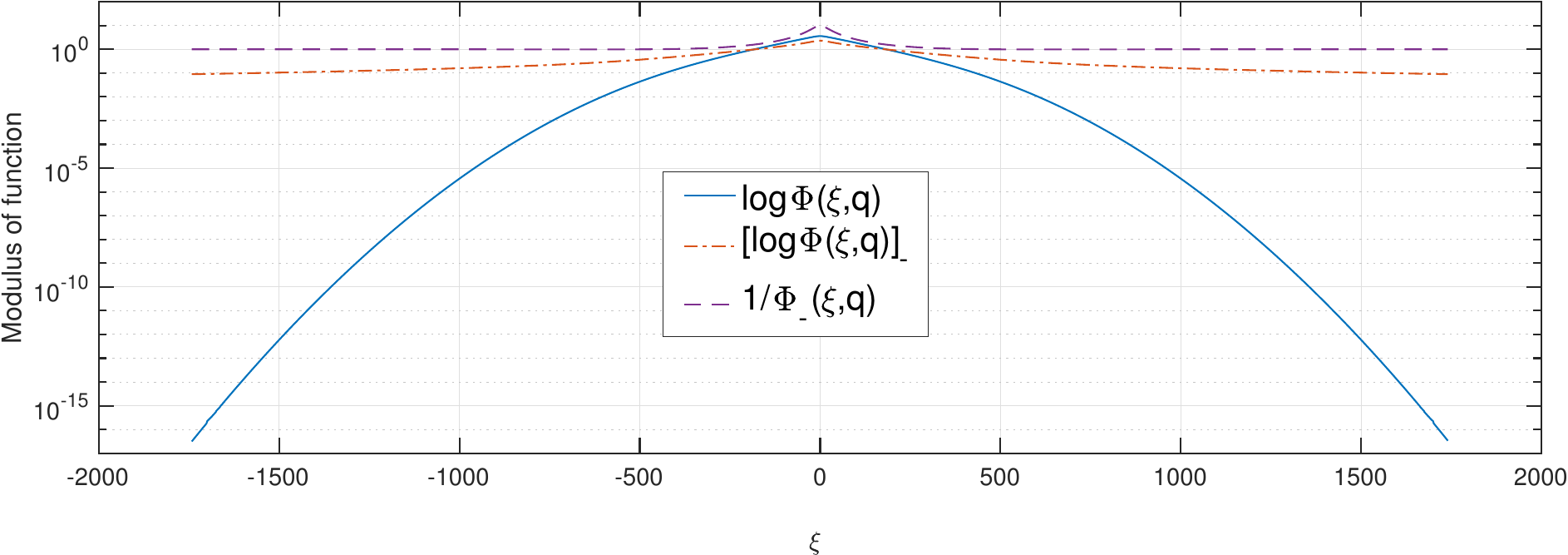}
\end{center}
\caption{Input and output functions for the factorisation of $\Phi(\xi,q)=1-q\Psi(\xi,\Delta t)$ with the Kou process for $q=\rho$. The plot shows how the decay of the function is changed by the decomposition. }
\label{fig:fact_input_output}
\end{figure}

The requirement to multiply the output of the factorisation by the characteristic function is due to its shape in the Fourier domain as this will influence the error convergence of the subsequent step.
Figure~\ref{fig:fact_input_output} shows that the function flattens out at high values of $|\xi|$ and asymptotically approaches $1$.
Therefore, if we were to input $\Phi_\pm(\xi,q)$ directly to the Hilbert transform in the decomposition step then we would not be able to bound the truncation error using Feng and Linetsky's error limit for exponentially bounded functions.

However, the last date is taken out of the FGM pricing scheme. This means that we multiply the function to be decomposed by the characteristic function. In the case of exponentially decaying characteristic functions, this restores the exponential decay of the function for high values of $\xi$ which again means that the truncation error of the discrete Hilbert transform is exponentially bounded. However, if the VG process is used then the input to the decomposition is only polynomially decaying and thus we again have polynomial error convergence for this stage.
\subsection{Double\,-barrier options with the unfiltered Spitzer identity}\label{sec:Errperf_Doub}
The original pricing procedure for double-barrier options shows polynomial convergence for all processes, even those whose characteristic function decays exponentially. The main difference between the pricing procedure for single and double-barrier options is the presence of the fixed-point algorithm and in this section we show how this causes the polynomial error convergence. As shown in Section~\ref{sec:Errperf_Sing}, with an exponentially decaying characteristic functions the factorisation has exponential error convergence. In addition we multiply the input to the fixed-point algorithm by the characteristic function, which means that it is exponentially bounded as $|\xi|\rightarrow\infty$.
Provided the input function to the first iteration of the fixed-point algorithm is exponentially bounded, the error on the output of the initial decomposition is exponentially bounded. However, the decomposition operation is equivalent to multiplying the function in the $x$ domain by either $\mathbf{1}_{\mathbb{R}_+}(x)$ or $\mathbf{1}_{\mathbb{R}_-}(x)$, which introduces a jump into the output functions. Due to the Gibbs phenomenon, this means that the output function from the decomposition decays as $O(1/\xi)$ as $|\xi|\rightarrow\infty$. The effect of this is that the input function to the second iteration of the fixed-point algorithm is no longer exponentially bounded and so, according to \cite{Stenger1993} and \cite{Feng2008}, the error from the truncation of the infinite sum in Eq.~(\ref{eq:HilbSincApprox}) to give Eq.~(\ref{eq:HilbSincApproxTrunc}) is no longer exponentially bounded. A bound for this error is
\begin{align}
|f_{\Delta\xi}(\xi)-f_{\Delta\xi,M}(\xi)| 
& <c_1\Delta\xi\sum_{k>M/2}\frac{1}{(k\Delta\xi)^2}\nonumber\\
& <c_2\int^{+\infty}_{M\Delta\xi/2}\frac{1}{\xi_k^2}d\xi_k
  <c_3\frac{1}{M\Delta\xi}.\label{eq:decompbound}
\end{align}
Therefore, using the fixed-point algorithm with more than one iteration means that the error is no longer exponentially bounded. The bound shown in Eq.~(\ref{eq:decompbound}) is $O(1/M)$. However, the error of the pricing procedure actually decays as $O(1/M^2)$; this better performance may be due to the alternating nature of the Fourier coefficients.
\subsection{Feng and Linetsky pricing method with the VG process}\label{sec:Errperf_FL}
The FL method is described in Eqs.~(\ref{eq:Hilbpricsing}) and (\ref{eq:Hilbpricdoub}), which show how the Hilbert transform is applied for each monitoring date. As explained in Section~\ref{sec:Errperf_Doub}, the application of the Hilbert transform introduces a discontinuity into the function in the log-price domain, therefore the Fourier coefficients on the output of the Hilbert transform will decay as $O(1/\xi)$ as $|\xi|\rightarrow\infty$. However, before the Hilbert transform is applied for the next monitoring date, the Fourier domain function is multiplied by the characteristic function of the underlying process. Therefore, as explained by \cite{Feng2008}, if the characteristic function is exponentially decaying, this will result in an exponentially convergent error. However, with polynomially decaying characteristic functions, such as that of the VG process, then a polynomially convergent error will be achieved.
\subsection{Error convergence with filtering on the sinc-based Hilbert transform}\label{sec:Errperf_filt}
The multiplication by a filter with exponentially decaying coefficients as $|\xi|\rightarrow\infty$ gives an exponentially convergent truncation error for the sinc-based discrete Hilbert transform compared with the non-truncated version. However, filtering distorts the function somewhat. The numerical results with the updated method are shown in Section~\ref{sec:results} and the prices calculated with the filtered version have been compared with the price calculated using the unfiltered FL method with the maximum grid size to confirm that any distortion error is less significant than the improvement in error convergence. Due to the error being influenced by these two opposing effects, we have not attempted to devise a tight error bound which closely matches the improvement in performance achieved in practice. It is often seen in the literature on the Gibbs phenomenon that the empirical results outstrip the calculated error bounds. For example, \cite{ruijter2015application} suggest that the faster convergence they see may be due to the alternating nature of the Fourier coefficients.
\section{Numerical results} \label{sec:results}
We performed numerical tests using the pricing schemes updated to include filtering, as described in Section~\ref{sec:Back}. The results for the FGM method for double-barrier options with exponentially decaying characteristic functions are presented in Section~\ref{sec:Res_exp}. Section~\ref{sec:Res_pol} contains results for all methods with the VG process. Details of the contract and the model parameters are included in Table~\ref{tab:Parasetup} in Appendix A. The numerical results were obtained using MATLAB R2016b running under OS X Yosemite on a 2015 Retina MacBook Pro with a 2.7GHz Intel Core i5 processor and 8GB of RAM.

\subsection{Results with exponentially decaying characteristic functions}\label{sec:Res_exp}
\begin{figure}[h]
\begin{center}
\includegraphics[width=\textwidth]{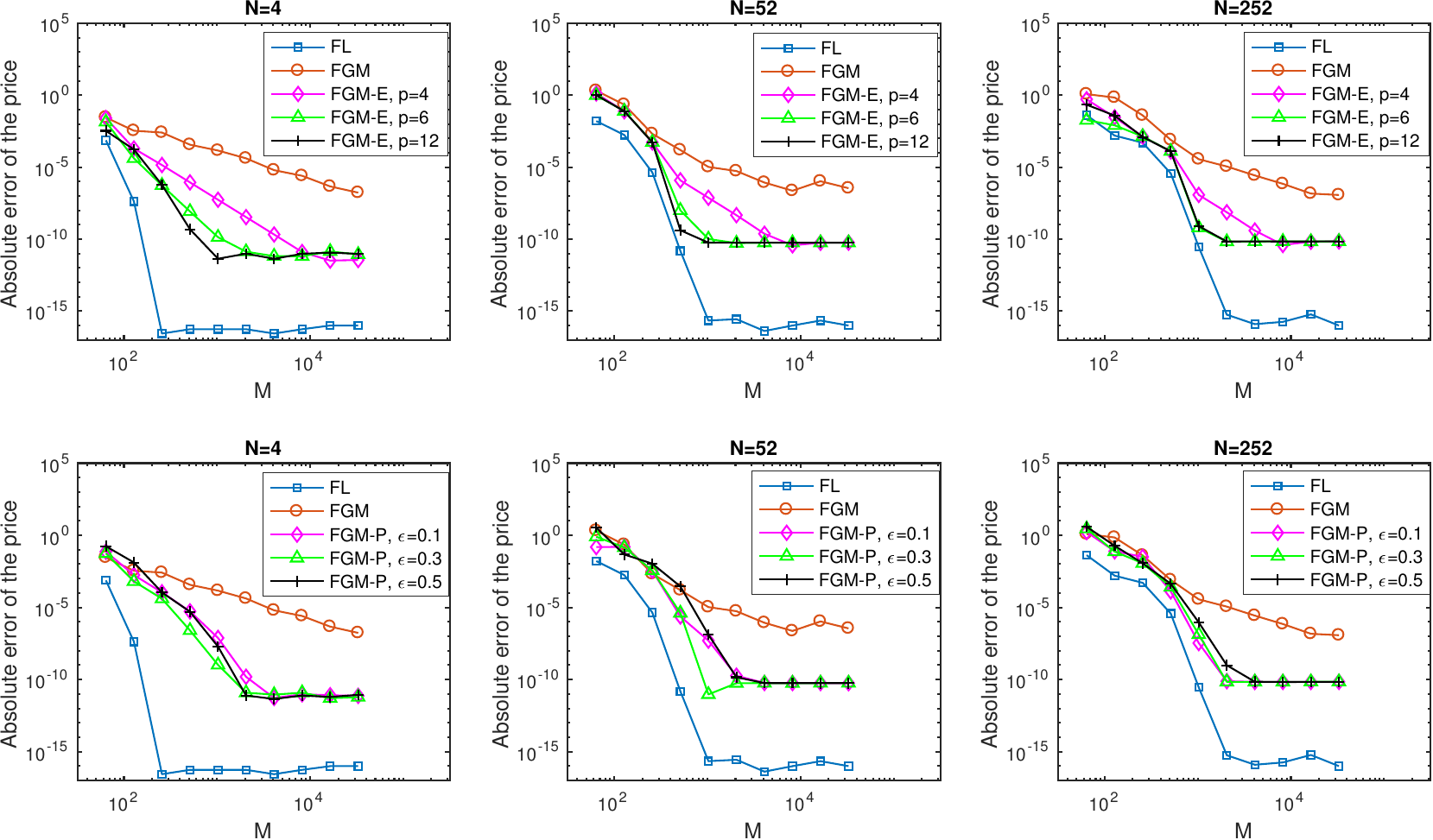}
\end{center}
\caption{Error vs.\ grid size $M$ for the Kou process and varying number of monitoring dates $N$. The filter improves the convergence of the FGM method from polynomial to exponential. The best results are obtained with an exponential filter of order $p=12$.}
\label{fig:N=all_U=1_15_L=0_85_KOU5}
\end{figure}

\begin{figure}[h]
\begin{center}
\includegraphics[width=\textwidth]{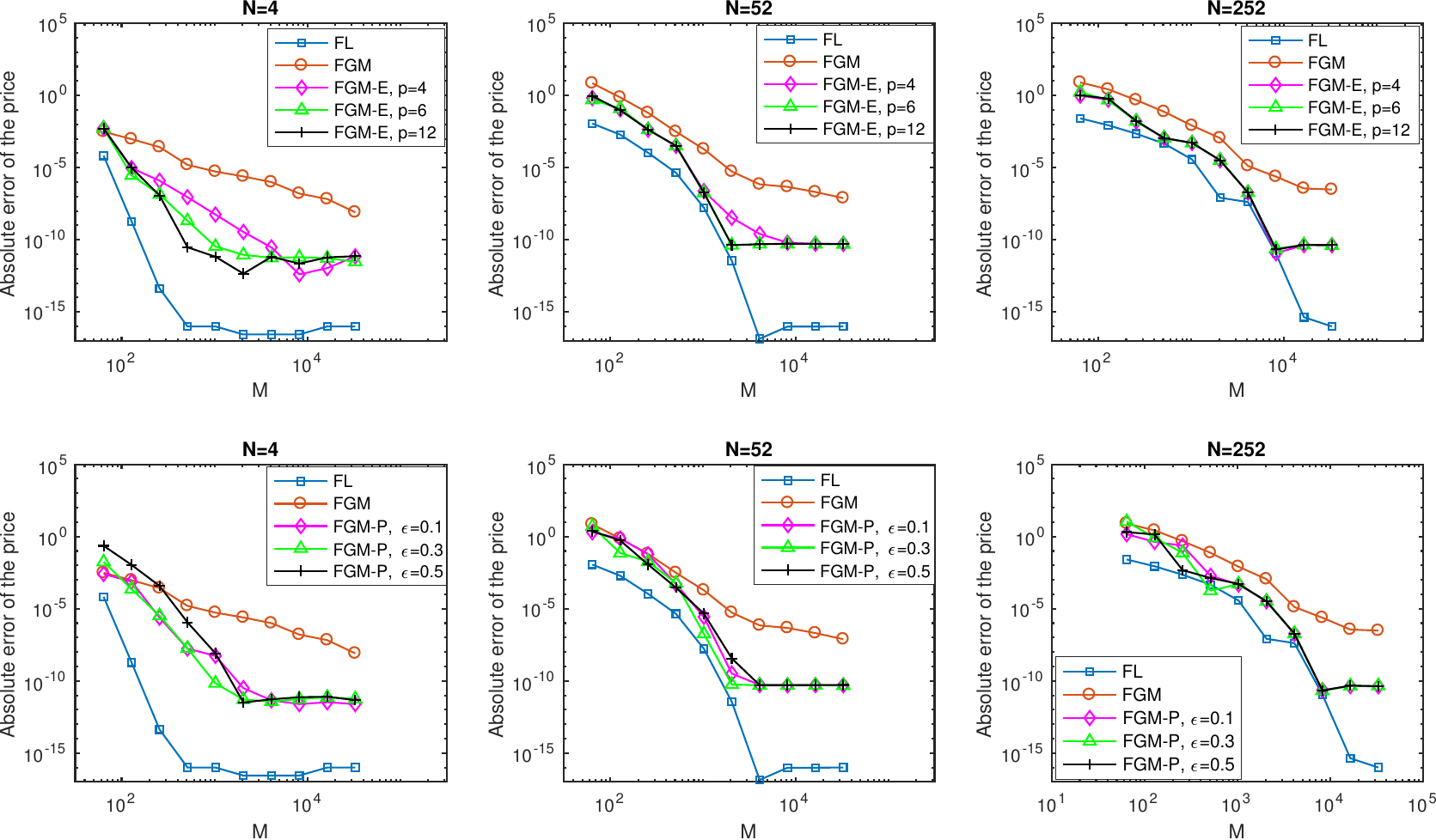}
\end{center}
\caption{Error vs.\ grid size $M$ with the NIG process and varying number of monitoring dates $N$. The filter improves the convergence of the FGM method from polynomial to exponential. The best results are obtained with an exponential filter of order $p=12$.}
\label{fig:N=all_U=1_15_L=0_85_NIG}
\end{figure}

We present results for the FGM method for double-barrier options with filtering included in the fixed-point algorithm as described in Section~\ref{sec:back_num_doub}. We examined the performance for both the Kou and NIG processes with $N=4$, 52 and 252.\footnote{The online supplementary material which accompanies this paper demonstrates the robustness of the pricing algorithm with regards to the variation of parameters.} The values of 52 and 252 represent weekly and daily monitoring over 1 year. Results with $N=4$ are presented in order to show the performance of the method with very few monitoring dates.
Figure~\ref{fig:N=all_U=1_15_L=0_85_KOU5} shows results for the Kou process and Figure~\ref{fig:N=all_U=1_15_L=0_85_NIG} shows results for the NIG process. The original FL and FGM methods are labelled ``FL'' and ``FGM''. The FGM method with filtering is labelled ``FGM-E, $p$=order'' for results with the exponential filter and ``FGM-P, $\epsilon$=parameter'' with the Planck taper. Comparing the results for all methods, we see that the FL method gives the best error convergence versus grid size. This is due to the error of the FGM method being limited by the performance of the inverse $z$-transform. Comparing the filtered FGM methods, the exponential filter gives better results but the Planck taper is less sensitive to variations in the filter shape. The best results were achieved with an exponential filter of order $p=12$.

\begin{figure}
\begin{center}
\includegraphics[width=\textwidth]{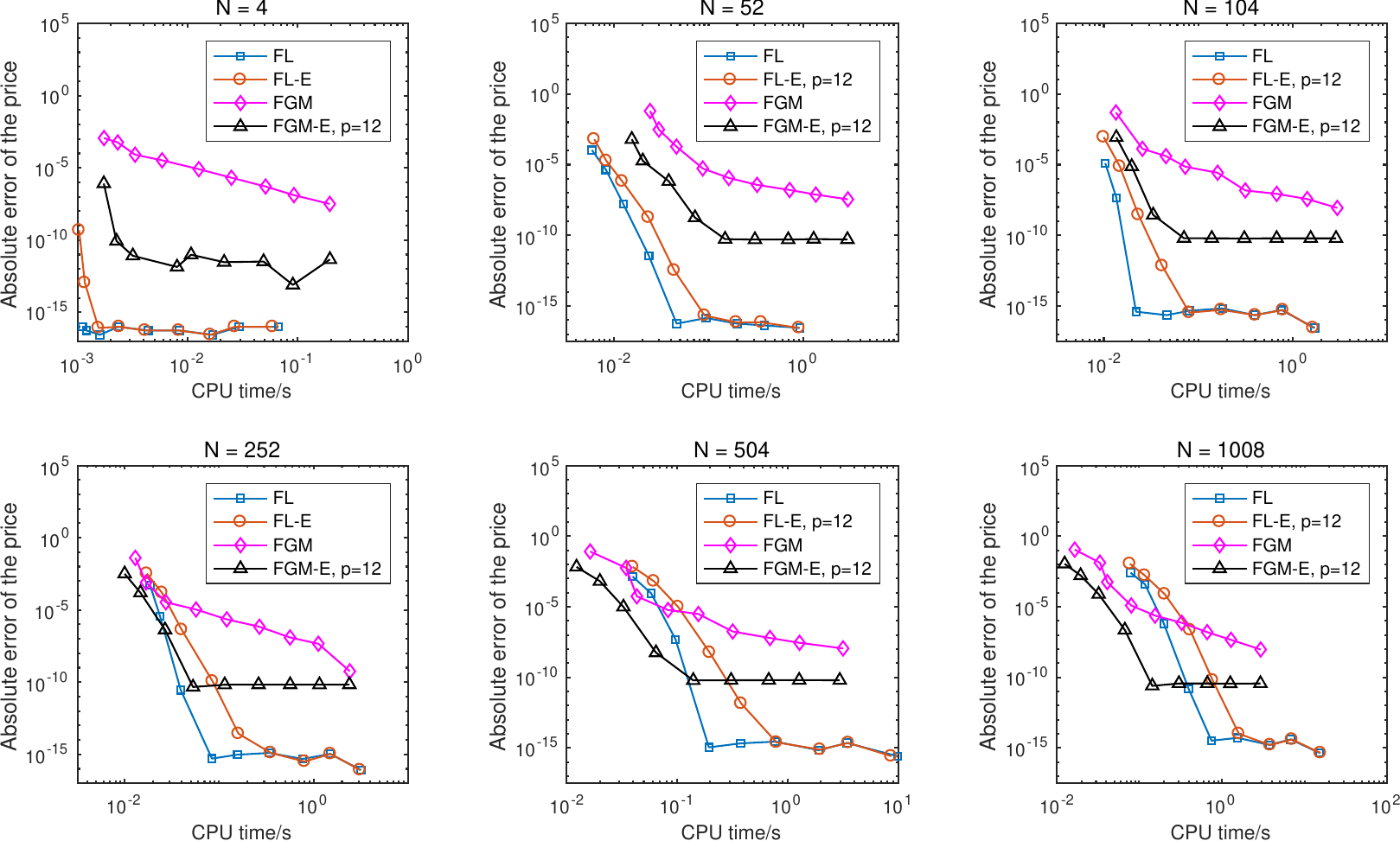}
\end{center}
\caption{Error vs.\ CPU time for a double-barrier option with the Kou process and varying numbers of monitoring dates $N$. The filter improves the FGM method for all $N$; FGM-E is the fastest method for an error of $10^{-8}$ with $N>252$.}
\label{fig:N=all_U=1_15_L=0_85_Kou_CPU}
\end{figure}

\begin{figure}
\begin{center}
\includegraphics[width=\textwidth]{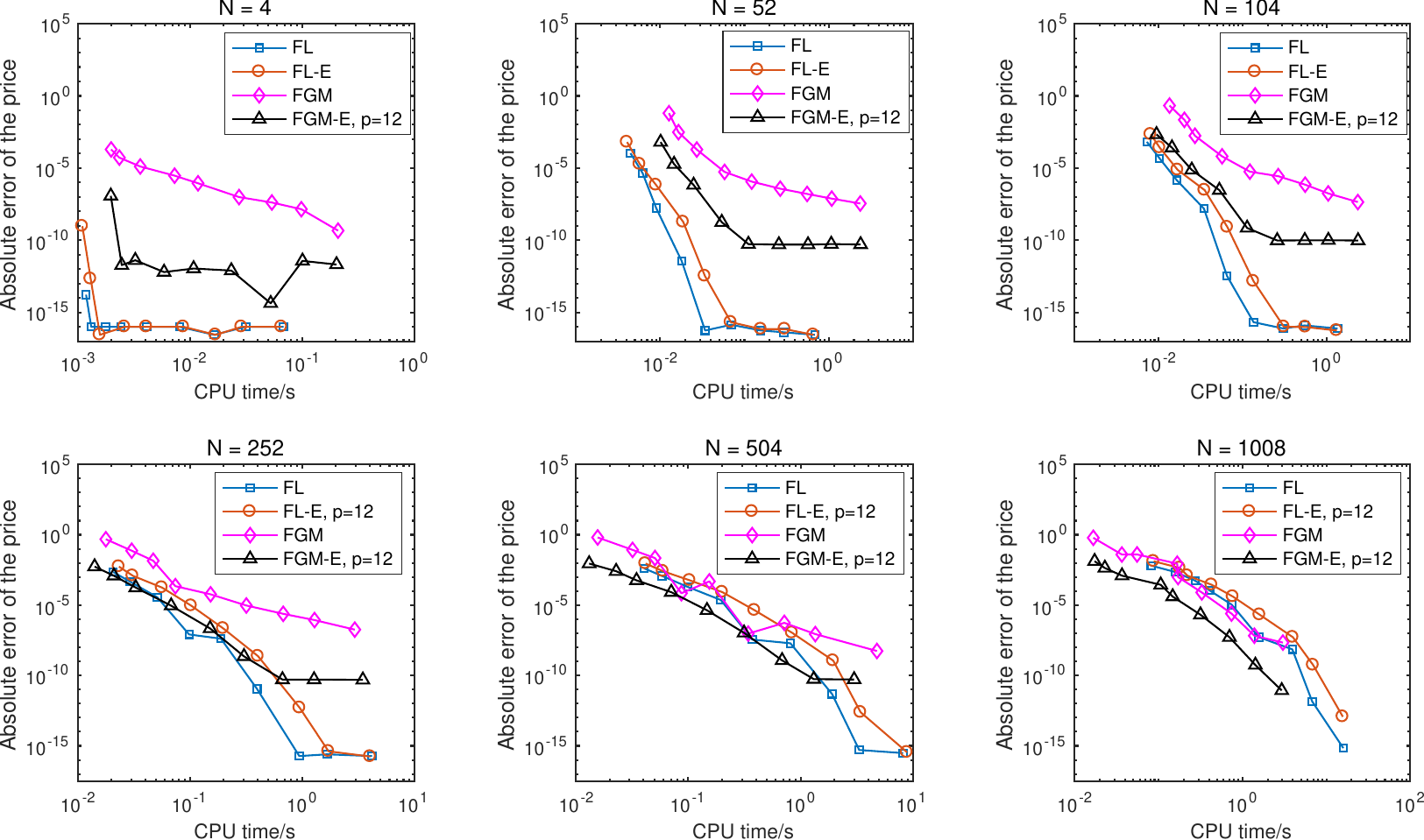}
\end{center}
\caption{Error vs.\ CPU time for a double-barrier option with the NIG process and varying numbers of monitoring dates $N$. The filter improves the FGM method for all $N$; FGM-E is the fastest method for an error of $10^{-8}$ with $N\geq504$.}
\label{fig:N=all_U=1_15_L=0_85_NIG_CPU}
\end{figure}

Tables~\ref{tab:Kouiterations} and \ref{tab:NIGiterations} present the number of iterations and the computational time for a range of dates. The results demonstrate that as the number of dates increases, the number of iterations and computational time either does not increase, or minimally increases, and thus confirm that the computational time is independent of the number of monitoring dates.
Figures~\ref{fig:N=all_U=1_15_L=0_85_KOU5} and \ref{fig:N=all_U=1_15_L=0_85_NIG} show how the convergence of the numerical techniques changes with the grid size and Figures~\ref{fig:N=all_U=1_15_L=0_85_Kou_CPU} and \ref{fig:N=all_U=1_15_L=0_85_NIG_CPU} show how the convergence behaviour corresponds to computational time with an exponential filter of order 12.

The inclusion of a filter in the FGM method produces a large improvement compared to the unfiltered method due to an increase in accuracy of the calculation, as shown in Figures~\ref{fig:N=all_U=1_15_L=0_85_KOU5} and \ref{fig:N=all_U=1_15_L=0_85_NIG}. However, this also improves the algorithm computationally as it now reaches the required accuracy in a smaller number of iterations than the original FGM method. Despite this improvement, for low numbers of monitoring dates the FL method shows the best performance. However, for 252 monitoring dates, the filtered FGM method performs around the same as the FL method for errors greater than $10^{-10}$; for higher number of dates, the filtered FGM method shows the best performance for errors greater than $10^{-10}$. Including the filter in the FL method produces a result with very slightly worse absolute error convergence but which still retains exponential convergence, see Figures~\ref{fig:N=all_U=1_15_L=0_85_Kou_CPU} and \ref{fig:N=all_U=1_15_L=0_85_NIG_CPU}. We can relate this to the error discussion in Section~\ref{sec:Errperf_filt}: the filter causes a slight distortion which degrades the absolute error convergence, but there is no improvement to be gained in the rate of convergence as the unfiltered method already achieves exponential convergence.

\begin{table}[h]
\centering
\begin{tabular}{rcccccc}
\hline
\hline
Dates & Tolerance & $M$ & Average iterations & Price & Error & CPU time \\
\hline
  4 & E-8  & 1024 & 2.000 & 0.00721968941 & 4.12E-14 & 5.63E-03 \\
 52 & E-8  & 1024 & 2.000 & 0.00518403635 & 3.07E-13 & 3.81E-02 \\
104 & E-8  & 1024 & 2.000 & 0.00490517113 & 5.54E-13 & 3.99E-02 \\
252 & E-8  & 1024 & 2.000 & 0.00465711572 & 4.29E-12 & 3.72E-02 \\
504 & E-8  & 1024 & 2.000 & 0.00452396360 & 4.31E-09 & 3.80E-02 \\
\hline
  4 & E-10 & 1024 & 2.000 & 0.00721968941 & 4.12E-14 & 1.82E-02 \\
 52 & E-10 & 1024 & 2.000 & 0.00518403635 & 3.07E-13 & 3.50E-02 \\
104 & E-10 & 1024 & 2.091 & 0.00490517113 & 5.62E-13 & 3.88E-02 \\
252 & E-10 & 1024 & 2.121 & 0.00465711572 & 4.31E-12 & 3.71E-02 \\
504 & E-10 & 1024 & 2.152 & 0.00452396360 & 4.31E-09 & 3.90E-02 \\		
\hline
\hline
\end{tabular}
\caption{Results for the Kou process with the fixed-point algorithm tolerance set to $10^{-8}$ and $10^{-10}$.}
\label{tab:Kouiterations}
\end{table}

\begin{table}[h]
\centering
\begin{tabular}{rcccccc}
\hline
\hline
Dates & Tolerance & $M$ & Average iterations & Price & Error & CPU time \\
\hline
  4 & E-8  & 1024 & 2.000 & 0.00545479385 & 2.38E-13 & 1.50E-02 \\
 52 & E-8  & 1024 & 2.000 & 0.00359559460 & 5.07E-13 & 8.57E-02 \\
104 & E-8  & 1024 & 2.000 & 0.00341651334 & 5.92E-10 & 8.58E-02 \\
252 & E-8  & 1024 & 2.091 & 0.00328484367 & 3.15E-07 & 9.63E-02 \\
504 & E-8  & 1024 & 2.182 & 0.00322814330 & 6.84E-07 & 9.34E-02 \\
\hline
  4 & E-10 & 4096 & 2.000 & 0.00545479385 & 7.17E-14 & 1.45E-02 \\
 52 & E-10 & 4096 & 2.242 & 0.00359559460 & 6.70E-13 & 2.20E-01 \\
104 & E-10 & 4096 & 2.303 & 0.00341651275 & 3.80E-13 & 2.15E-01 \\
252 & E-10 & 4096 & 2.364 & 0.00328453104 & 2.33E-09 & 2.08E-01 \\
504 & E-10 & 4096 & 2.485 & 0.00322753427 & 7.53E-08 & 2.21E-01 \\		
\hline
\hline
\end{tabular}
\caption{Results for the NIG process with the fixed-point algorithm tolerance set to $10^{-8}$ and $10^{-10}$.}
\label{tab:NIGiterations}
\end{table}

\subsection{Polynomially decaying characteristic functions}\label{sec:Res_pol}
\begin{figure}[h]
\begin{center}
\includegraphics[width=\textwidth]{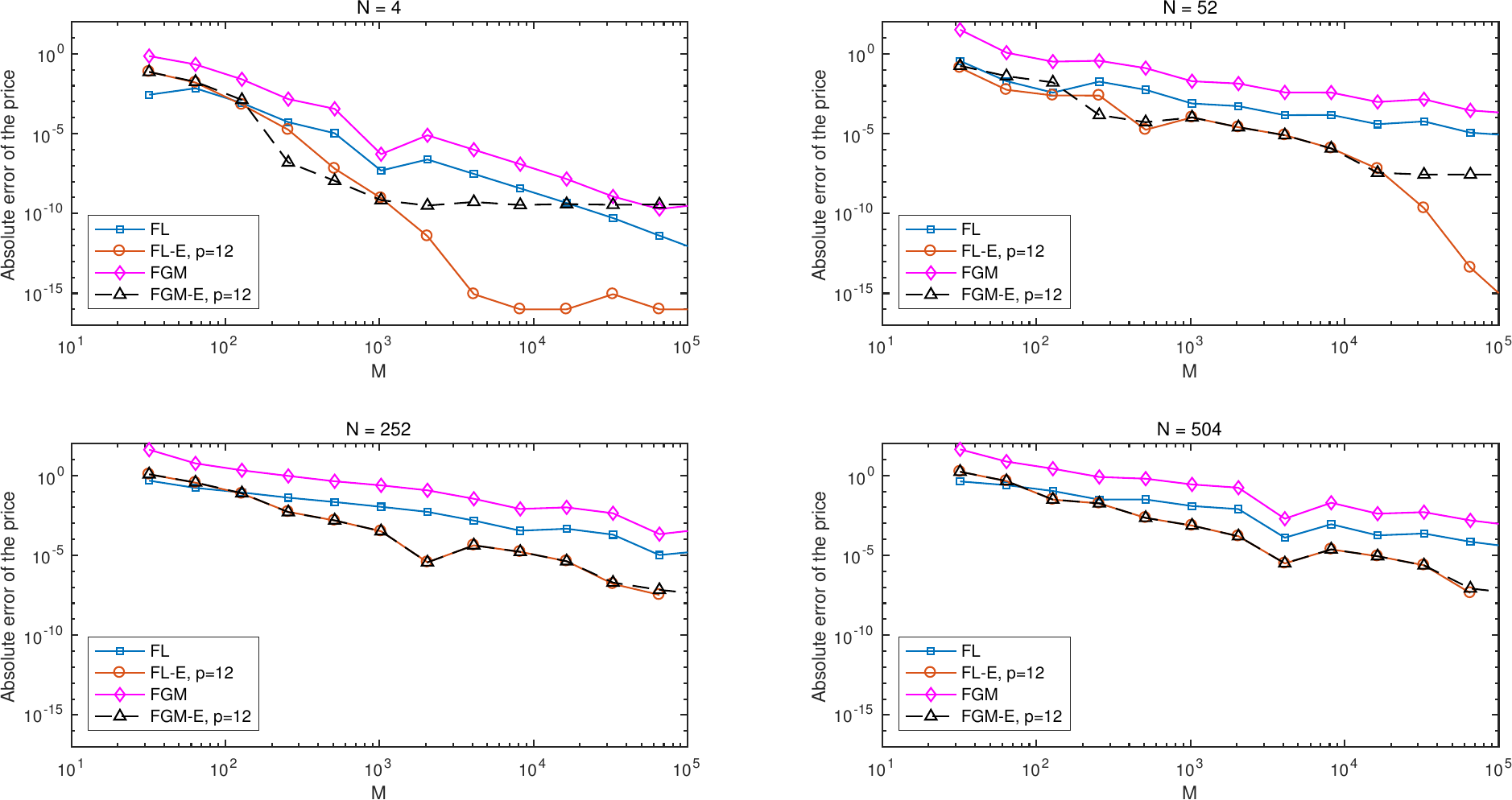}
\end{center}
\caption{Error vs.\ grid size $M$ for a single-barrier down-and-out option with the VG process and varying numbers of monitoring dates $N$. The filter improves both the FGM and FL methods, with the FL-E method performing best at low numbers of dates.}
\label{fig:N=all_U=trunc_L=0_85_VG}
\end{figure}

\begin{figure}
\begin{center}
\includegraphics[width=\textwidth]{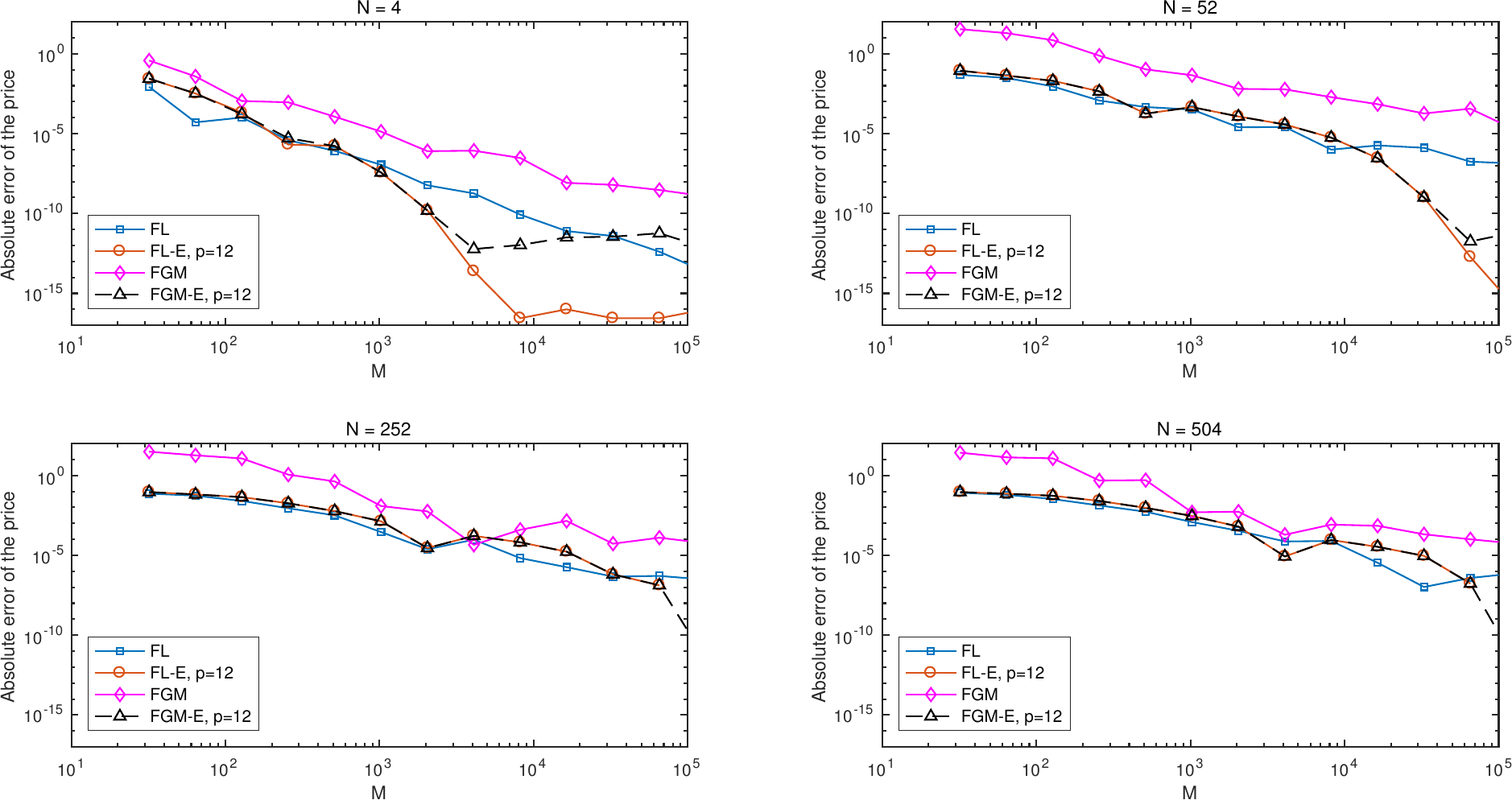}
\caption{Error vs.\ grid size $M$ for a double-barrier option with the VG process and varying numbers of monitoring dates $N$. The filter improves the FGM methods for all numbers of monitoring dates and the FL method for low numbers of dates.}
\label{fig:N=all_U=1_15_L=0_85_VG}
\end{center}
\end{figure}

We present results for the FL and FGM methods for a process with a polynomially decaying characteristic function, i.e.\ the VG process.
Figures~\ref{fig:N=all_U=trunc_L=0_85_VG} and \ref{fig:N=all_U=1_15_L=0_85_VG} show the results of tests for single and double-barrier options where we have applied exponential filtering as described in Section~\ref{sec:Back_num}.

\begin{figure}
\begin{center}
\includegraphics[width=\textwidth]{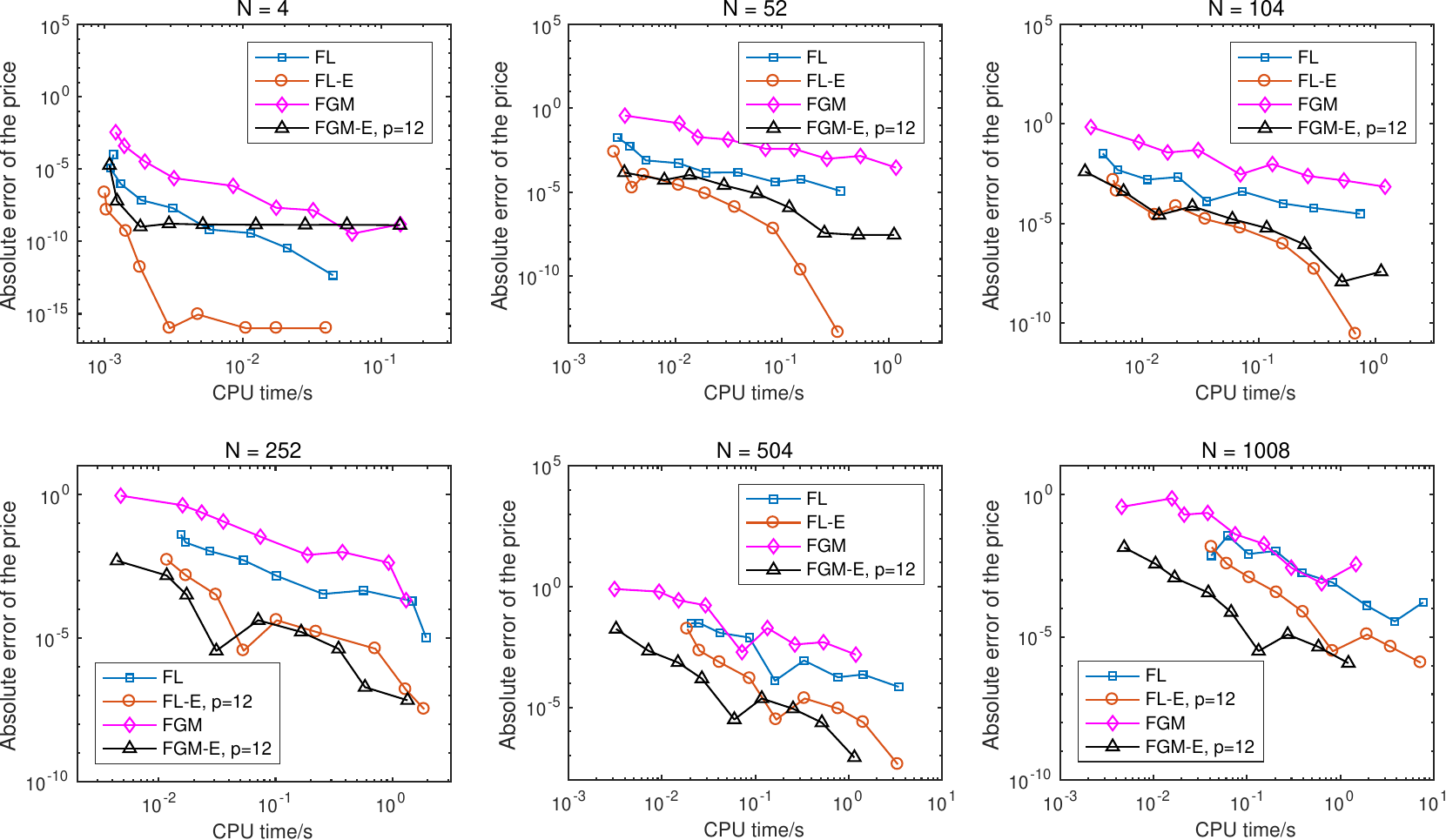}
\end{center}
\caption{Error vs.\ CPU time for a single-barrier option with the VG process and varying numbers of monitoring dates $N$. The best performance of the new filtered methods, FL-E, $p=12$, and FGM-E, $p=12$, either equals or exceeds the performance of the existing methods over all numbers of dates.}
\label{fig:N=all_U=trunc_U_L=0_85_VG_CPU}
\end{figure}

\begin{figure}
\begin{center}
\includegraphics[width=\textwidth]{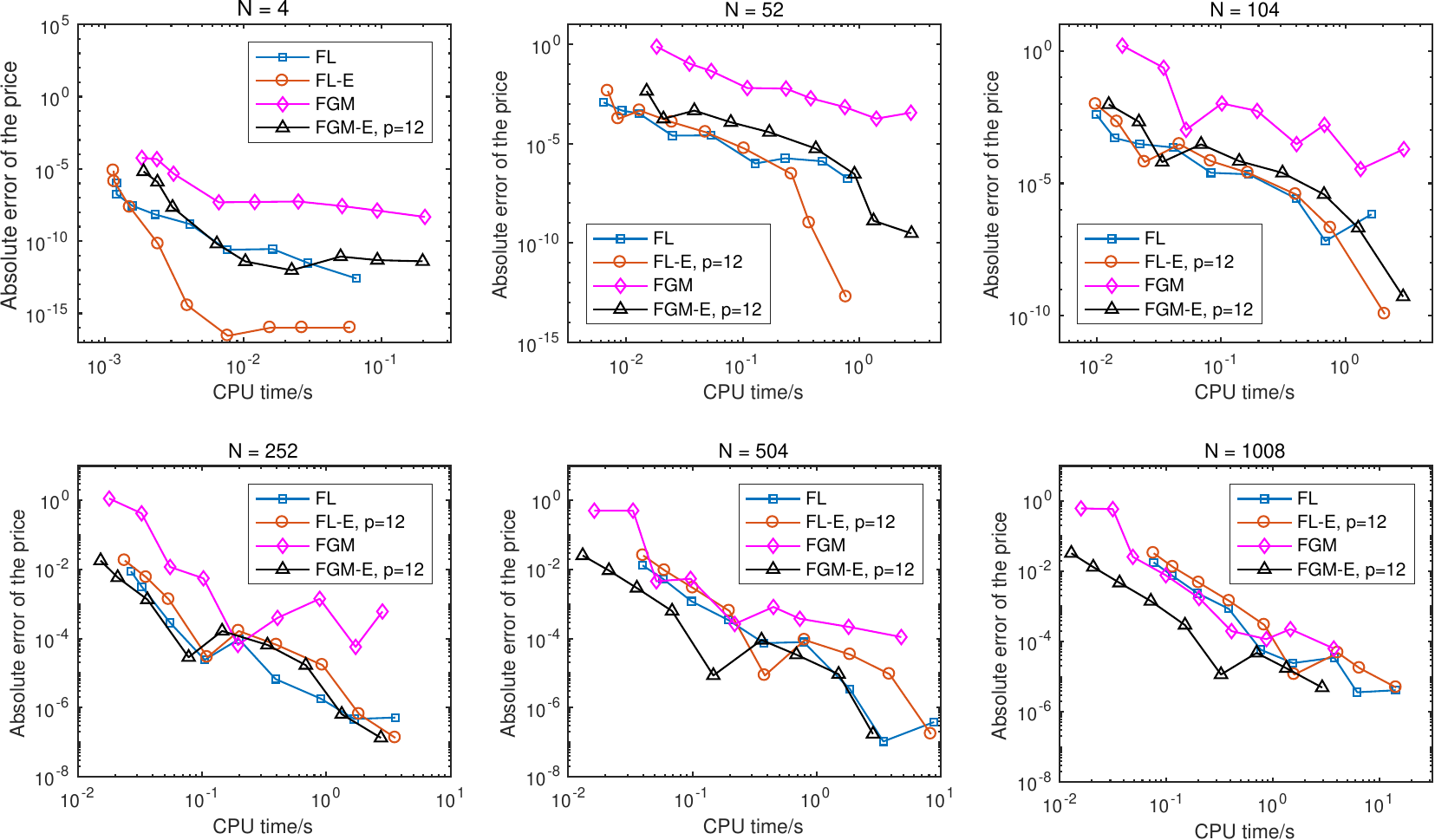}
\end{center}
\caption{Error vs.\ CPU time for a double-barrier option with the VG process and varying numbers of monitoring dates $N$. The best performance of the new filtered methods, FL-E, $p=12$ and FGM-E, $p=12$, either equals or exceeds the performance of the existing methods over all numbers of dates.}
\label{fig:N=all_U=1_15_L=0_85_VG_CPU}
\end{figure}

The performance for a low number of dates shows a good improvement with the addition of filtering for both the FGM and FL methods. This demonstrates that the performance of the sinc-based discrete Hilbert transform of polynomially decaying functions can be improved even when the polynomial decay is a true representation of the function shape and not simply an artefact of the fixed-point algorithm as was the case in Section~\ref{sec:Res_exp}. For a higher number of dates, the error convergence vs grid size for the FGM method is improved so that it is the same as the FL method with or without filtering. This is a significant improvement as the FGM method has the advantage over the FL method that its computation time beyond a small threshold is independent of the number of dates, unlike the linear increase of the FL method.
This is demonstrated by the results shown in Figures~\ref{fig:N=all_U=trunc_U_L=0_85_VG_CPU} and \ref{fig:N=all_U=1_15_L=0_85_VG_CPU}. The filtered methods show the best performance for all dates; filtered FL is the best performing method for low numbers of monitoring dates and filtered FGM is the best performing method for higher numbers of dates.
\FloatBarrier

\subsection{Summary of results}
Table~\ref{tab:BestMeth} shows a summary of the best performing methods in terms of CPU time for different processes and types of options.
\begin{table}[h]
\begin{center}
\begin{tabular}{r |c |c c c}
\hline
\hline
& Single barrier &\multicolumn{3}{c}{Double barrier}\\
\hline
Dates & VG & Kou & NIG & VG\\
\hline
   4 & \textbf{FL-E} & \textit{FL} &  \textit{FL} & \textbf{FL-E}\\
  52 & \textbf{FL-E} & \textit{FL} & \textit{FL} & \textbf{FL-E}\\
 104 & \textbf{FL-E, FGM-E} & \textit{FL} & \textit{FL} & \textbf{FL-E}\\
 252 & \textbf{FL-E, FGM-E} & \textit{\textbf{FGM-E, FL}} & \textit{\textbf{FGM-E, FL}} & \textit{\textbf{FGM-E, FL-E, FL}}\\
 504 & \textbf{FGM-E$^*$} & \textbf{FGM-E} & \textbf{FGM-E} & \textbf{FGM-E$^*$}\\
1008 & \textbf{FGM-E$^*$ }& \textbf{FGM-E} & \textbf{FGM-E} & \textbf{FGM-E$^*$}\\
\hline
\hline
\end{tabular}
\end{center}
\caption{Quickest method for an error of $10^{-8}$. Due to the slower convergence of all methods with the VG process, entries marked with an asterisk show the quickest method for an error of $10^{-5}$.
\textbf{Bold}: a filtered method provides the best performance. \textit{\textbf{Bold italic}}: the performance of the filtered methods equals, but does not exceed, the best performance of an existing method. \textit{Italic}: the few cases where an existing method performs best.}
\label{tab:BestMeth}
\end{table}

\section{Conclusions}
In this article we showed that numerical methods for pricing derivatives based on the Hilbert transform computed with a sinc function expansion can be modified with the addition of spectral filters to improve their convergence. Furthermore, we expanded on the work by Stenger and Feng and Linetsky which showed how the shape of the function on the input to the Hilbert transform relates to the resultant error on the output of the Hilbert transform. We showed that due to the Gibbs phenomenon, an algorithm using successive Hilbert transforms will achieve polynomial convergence unless additional filtering is applied after the first Hilbert transform. Moreover, we demonstrated that simple spectral filters such as the exponential filter or the Planck taper are sufficient to improve performance so that exponential convergence can be achieved.
In addition we showed that the pricing schemes by Feng and Linetsky and Fusai et al., which have relatively poor performance with the VG process, even for single-barrier options, can also be improved by spectral filters.
This article directly concerns the pricing of barrier option pricing but the findings are relevant for any application which is related to L\'evy processes in the presence of barriers and requires the solution of the Wiener-Hopf or Fredholm equation.


\bibliographystyle{spbasic} 
\bibliography{DoubleBarrierPaper} 
\newpage
\section*{Appendix A}
Table~\ref{tab:Parasetup} contains all the parameters used for the numerical experiments which produced the results presented in Section~\ref{sec:results}.
\begin{table}[h]
\begin{center}
\begin{tabular}{lllr}
\hline\hline
& Description & Symbol & Value \\
\hline
\multirow{9}{*}{Option parameters}&Maturity & $T$& 1 year \\
&Initial spot price &$S_0$ & 1\\
&Strike &$K$ & 1.1\\
&Upper barrier (double-barrier) & $U$ & 1.15\\
&Upper barrier (down-and-out) & $U$ & $+\infty$\\
&Lower barrier & $L$ & 0.85\\
&Risk-free interest rate & $r$ & 0.05\\
&Dividend rate & $q$& 0.02\\
\hline
Model & $\Psi(\xi,t)$ & Symbol & Value\\
\hline
\multirow{3}{*}{NIG}&\multirow{3}{*}{$e^{-t\delta\left(\sqrt{\alpha^2-(\beta+i\xi)^2}+\sqrt{\alpha^2-\beta^2}\right)}$} & $\alpha$ & 15\\
& & $\beta$ & -5\\
& & $\delta$ & 0.5\\
\hline
\multirow{5}{*}{Kou}&\multirow{5}{*}{$e^{-t\left(\frac{\sigma^2\xi^2}{2}-\lambda\left(\frac{(1-p)\eta_2}{\eta_2+i\xi}+\frac{p\eta_1}{\eta_1-i\xi}-1\right)\right)}$} & $p$ & 0.3\\
& & $\lambda$ & 3\\
& & $\sigma$ & 0.1\\
& &$\eta_1$ & 40\\
& & $\eta_2$ & 12\\
\hline
\multirow{3}{*}{VG}&\multirow{3}{*}{$(1-i\nu\theta\xi+\nu\sigma^2\xi^2/2)^{-t/\nu}$} & $\theta$ & $1/9$\\
\noalign{\vskip 0.5mm}
& & $\sigma$ & $\sqrt{3}/9$\\
\noalign{\vskip 0.5mm}
& & $\nu$ & 1/4 \\
\hline\hline
\end{tabular}
\end{center}
\caption{Parameters for the numerical tests and processes used; $\Psi(\xi,t)$ is the characteristic function of the process that models the transition density of the log-price of the underlying asset.}
\label{tab:Parasetup}
\end{table}
\end{document}